\documentclass{article}
\usepackage{graphicx}

\author{Aron C. Wall\footnote{aroncwall@gmail.com}
\\ \textit{Department of Physics} \\ \textit{University of California, Santa Barbara}
\\ \textit{Santa Barbara, CA 93106, USA} }
\title{A discrete, unitary, causal theory of quantum gravity}

\date{\today}

\begin{document}

\maketitle

\begin{abstract}
A discrete model of Lorentzian quantum gravity is proposed.  The theory is completely background free, containing no reference to absolute space, time, or simultaneity.  The states at one slice of time are networks in which each vertex is labelled with two arrows, which point along an adjacent edge, or to the vertex itself.  The dynamics is specified by a set of unitary replacement rules, which causally propagate the local degrees of freedom.  The inner product between any two states is given by a sum over histories.  Assuming it converges (or can be Abel resummed), this inner product is proven to be hermitian and fully gauge-degenerate under spacetime diffeomorphisms.  At least for states with a finite past, the inner product is also positive.  This allows a Hilbert space of physical states to be constructed.
\newline\newline
PACS numbers: 04.60.-m, 04.60.Gw, 04.60.Nc
\end{abstract}

\newpage
\tableofcontents
\newpage

\section{Introduction}

This article will propose a class of quantum gravitational models describing a background-free, discrete spacetime, which evolves by means of local causal unitary evolution.  The models in this class will be referred to as DUCT's (Discrete Unitary Causal Theories).  The kinematics of the theory will be loosely inspired by loop quantum gravity, but the primary goal will be to produce consistent dynamics rather than to retain contact with the loop quantization of general relativity. 

\subsection{Motivating Remarks}\label{motive}

Supposing we think of the universe as a computer---what kind of computer would it be?  

\paragraph{Quantum}  First of all, the universe is a \emph{quantum} computer, since the world displays quantum mechanical interference between possible histories.  The dynamics of a quantum system are described by unitary operators acting on a Hilbert space of states.  This is in contrast to a classical system, whose dynamics are described by functions acting on sets of states.

\paragraph{Background Free}  Secondly, general relativity leads us to believe that the universe is a computer in which space and time are themselves part of the dynamical state.  Systems with this property are called \emph{background free} (also known as \emph{background independent}).  A background-free computer is not like the computers which exist on our desktops, since these computers exist in a stable spacetime which they do not affect or control, and upon which they depend.\footnote{Actually, though, this difference is not as profound as it seems at first sight, since even though computer programs can depend on physical space and time, for the most part (other than clock applications) we don't want them to.  Although we want our programs to evaluate quickly, usually the end result should not depend on how much time each process took, or where the program was stored in the computer.}

A background-free theory should have at least the following properties:
\begin{enumerate}
\item The kinematics of the theory (which describe space at one time) should be most naturally expressed in a way which is invariant under spatial diffeomorphims.

\item The dynamics of the theory (describing how things change with time) should be a pure gauge symmetry.  When quantizing  canonical general relativity, this is equivalent to imposing the Wheeler-de Witt Hamiltonian constraint equation 
$H \Psi = 0$.  Since most kinematic states do not have this property, there is a dramatic reduction in the degrees of freedom of the system as a result of this constraint.

\item There should be no absolute notion of simultaneity.  In other words, there is no invariant meaning to ``the same time, but a different place''.  In canonical general relativity, this comes about because the Hamiltonian $H$ depends on the lapse, which is an arbitrary function of position.  There are many different possible ways to foliate the same spacetime \cite{kuchar72}.  (As Wheeler put it, time is ``many-fingered'' \cite{MTW}.)
\end{enumerate}

\paragraph{Discrete}  A third and more contentious question is whether the universe is a digital or analog computer.  A digital spacetime would be made out of discrete components which evolve forward by discrete time-steps.  An analog computer might have continuous parts, as in field theory.  There is some evidence that quantum gravity should be digital, from the finiteness of black hole entropy in generalized thermodynamics \cite{finiteS} (based on the idea that the black hole entropy should count the entropy of some set of microscopic degrees of freedom associated with the horizon \cite{micro}.)  Also, in loop quantum gravity, area and volume operators have discrete eigenvalues \cite{RS95}.

However, it is very difficult to see how Lorentz invariance could act as a symmetry on a discrete spacetime \cite{lorentz, BHS09},\footnote{except in the causal set approach, discussed below.} although Lorentz invariance seems to be necessary for the thermality of Hawking radiation \cite{thermal}, and the validity of the generalized second law of horizon thermodynamics \cite{EFJW07}.  Furthermore, the asymptotic safety scenario \cite{asympt} suggests the possibility of a continuum theory of quantum gravity.  There also seems to be no shortest distance in string theory (even though in some respects the string length acts as a UV cutoff on the theory) although it is hard to say what a background-free formulation of string theory might look like.

Since the theoretical evidence cuts both ways and there are no experiments, it would be unwise to be excessively dogmatic about the issue.  Nevertheless, for the remainder of this article the digital model will be assumed as a working hypothesis.  (Since the discreteness is assumed to be fundamental rather than a lattice approximation, I will assume that this discrete theory is exactly background-free rather than approximately so.)

\begin{figure}[ht]
\centering
\includegraphics[width=.39\textwidth]{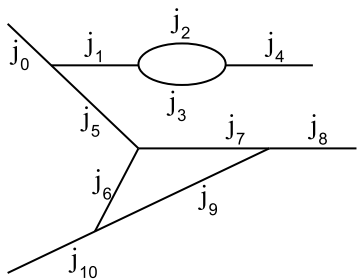}
\caption{\footnotesize In loop quantum gravity, kinematic states are described by superpositions of networks.  Here is an example of a trivalent spin network.  Each edge is labelled by a representation of SU(2) (i.e. a positive half-integer ``spin'' $j_a$).  There are constraints on the spins that can meet at any vertex.  In this article, kinematic states will also be assumed to be a labelled network.  However, the kinds of labels used will differ from loop quantum gravity.}\label{spinnet}
\end{figure}

Thus, the kinematics of a DUCT will be described by some discrete topological structure.  Inspired by loop quantum gravity (Fig. \ref{spinnet}), I will suppose that space at one time can be described by quantum superpositions of labelled networks.  Unlike loop quantum gravity, however, the goal will not be to quantize a pre-existing classical theory, but rather to construct fully background-free quantum causal spacetime dynamics, using whatever structures are convenient for doing so.

\begin{figure}[ht]
\centering
\includegraphics[width=.9\textwidth]{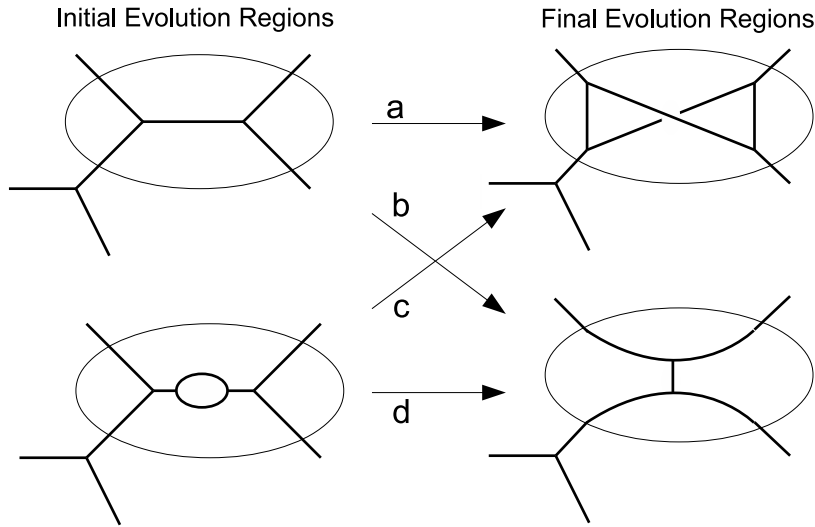}
\caption{ \footnotesize A local replacement rule.  A network is evolved to the future by performing a substitution in a single region (whose boundary is shown as an ellipse).  The laws of the DUCT might designate that the two possible states shown on the left are ``initial evolution regions'' (IER's).  Any such IER may be evolved forwards in time to ``final evolution regions'' (FER's), which have the same boundary (in this case, four edges).  If these are the only IER's and FER's with this boundary, then the transition amplitudes are given by a 2x2 matrix of transition amplitudes: $U = [a,\,b;\,c,\,d]$.  In order to conserve probability, it is necessary that $U$ be unitary.}\label{replace}
\end{figure}

The dynamics of a DUCT consist of discrete steps in which the labelled network is modified by means of ``local replacement rules'' (to be described in more detail below) in which one portion of the network is replaced by another in some local region.  These local replacement rules will be described by invertible unitary transformations $U$ defining the quantum mechanical transition amplitudes to go from one network state to another (Fig. \ref{replace}).  Let a region that can be evolved forwards in time be called an ``initial evolution region'' (IER), and let a region that can be evolved backwards in time be called a ``final evolution region'' (FER).  Each unitary transformation $U$ will take the form of an operator going from a Hilbert space of IER's to a Hilbert space of FER's.

Those familiar with canonical quantum gravity should be aware of two differences with the formalism used here:

First, in a theory with discrete time, there is no Hamiltonian operator $H$.  That is because the Hamiltonian is defined as the generator of an infinitesimal time translation, and there is no such thing as an infinitesimal time translation if time proceeds by discrete steps only.  Instead, the collection of invertible unitary transformations $U$ will play the role of the Hamiltonian operator.  There will be one $U$ matrix for every kind of $\mathrm{IER} \to \mathrm{FER}$ transition permitted by the theory.  (Since the domain and the target of $U$ are two distinct vector spaces, it does not make sense to talk about eigenvalues of $U$.)

Second, since time evolution is a gauge symmetry, $U$ will be regarded as expressing an exact physical equivalence between the initial and final states related by the unitary transformation.  In other words, the physical Hilbert space will be constructed by modding out all vectors of the form $U \Psi - \Psi$, where $\Psi$ is any state in the domain of $U$.  This is different from the usual picture in which the physical Hilbert space is realized by means of imposing the Wheeler-de Witt equation as a constraint.  However, the two pictures are equivalent.

One can go from one picture to the other simply by taking the dual of the vector space of states.  Let $\mathcal{H}_\mathrm{kin}$ be the vector space of kinematic states, and let $\mathcal{H}_\mathrm{phys}$ be the subspace satisfying some dynamical constraint $C \psi = 0$.  Then the dual physical state space $\mathcal{H}^*_\mathrm{phys}$ is a quotient space of the dual kinematic state space $\mathcal{H}^*_\mathrm{kin}$.  This quotient is obtained by modding out by all dual vectors $\chi$ satisfying $\chi \cdot \psi$ = 0 for all $\psi$ in the kernel of the constraint $C$.

The dual picture is convenient because it is not necessary to solve a complicated constraint in order to refer to a physical state.  Any kinematic state is also a dynamical state vector (although it may end up being the zero vector).  This article will use the dual picture exclusively.

\paragraph{Causal}  In addition to all this, the universe is a parallel processing computer.  In any relativistic theory such as general relativity or quantum field theory, there is a maximum speed of information propagation.  Consequently, spacelike separated objects can be evolved forwards in time independently of one another.  This is closely related to the ``many-fingered time'' aspect of background independence mentioned above.

Attempts have been made to construct causal theories of quantum gravity by imposing a causal structure (i.e. a partial ordering) on the elements of discrete spacetime histories.  For example, Markopoulou and Smolin \cite{MS97} construct a model in which a spin network evolves forwards in time by means of rules which causally propagate the spin information of the network.  In this model, time evolution rule is stated as a rule for evolving an entire \emph{global} spin-network $\Gamma_0$ forwards in time everywhere to a ``successor'' spin-network $\Gamma_1$.  This means that unlike general relativity, the theory is not fully background free, because it is formulated in a way which relies on an absolute notion of simultaneity.  Thus there is no local time-translation symmetry, no many-fingered time, and no Hamiltonian constraint.

Markopoulou \cite{markopoulou97} generalizes the preceding model by permitting arbitrary Pachner moves to act on spin-networks.  She also suggests generalizing the model to the ``non-maximal'' case in which not all spin-network vertices are evolved forwards in time on each time-step.  The article claims that these modified models can be used to construct spacetimes which have a varying spacetime foliation.  This is true in the sense that each individual spacetime has a causal structure which can be foliated in multiple ways by spacelike surfaces.  But it is not clear that the model is still unitary and causal once one sums over different spacetime histories.  Hence it has not been shown that the dynamics are truly invariant under reparameterizing time as a function of space.  As the article states, ``implementing causality now involves finding the correct choices of these amplitudes, which at this stage is an unsolved problem''.  See Refs. \cite{cfoam} for additional attempts to impose a causal structure for spin foam models.

Similarly, the ``causal dynamical triangulations'' approach of Ambjorn and Loll \cite{AL98} requires histories to be causal in the sense that each classical history admits a causal structure.  However, despite claims that this formalism is background free, the theory has built into it a preferred notion of spatial simultaneity.  Unlike general relativity, their construction privileges one particular foliation of spacetime into spatial slices.  Consequently, it does not qualify as background free in the sense used here.\footnote{Conceivably, this background structure might decouple from physical degrees of freedom after taking the continuum limit.  However, it seems to me that general relativity is more likely to emerge from a discrete theory which has full diffeomorphism invariance built in at the quantum level.}

In the ``causal set'' approach to quantum gravity \cite{csets}, spacetime is assumed to be nothing but a partially ordered set of events.  Causal sets have \emph{only} a discrete causal structure, which despite its discreteness appears to be compatible with Lorentz invariance \cite{BHS09}.  Although it is possible to define causal quantum theories living on fixed causal sets \cite{markopoulou99}, there does not yet exist a fully background-free quantum dynamics \emph{of} the causal set.  Furthermore, since the notion of an ``initial data slice'' appears to be absent in this approach, it may be difficult to reconstruct notions of states and observables.

For the reasons stated above, none of the above models can yet be regarded as fully satisfactory implementations of quantum causality.  Indeed, there are some important obstacles to constructing a truly causal discrete theory of spacetime.  The reason I have been at pains to emphasize the computational model of quantum gravity, is that these obstacles are already well-known and have names in the theory of parallel computation!  In computer science, these obstacles go by the names of \emph{race conditions} and \emph{deadlock}.

A computer with a parallel processor can run algorithms more quickly by splitting the work up into pieces and assigning them to independent processes called ``threads''.  However, problems can arise when the threads fail to be truly independent of each other.  A ``race condition'' occurs when two threads require conflicting access to the same computing resource---such as a bit of information $b$ stored in a computer.  For example, if thread A executes an instruction to read whether $b$ is 0 or 1, and thread B executes an instruction to toggle the bit from 0 to 1, then it matters which thread happens to execute its instruction first.  This is regarded as undesirable because the goal of an algorithm is to calculate a specific answer in a reliable fashion.  The answer should not depend on which of the two purportedly independent threads happens to execute first.

\begin{figure}[ht]
\centering
\includegraphics[width=.9\textwidth]{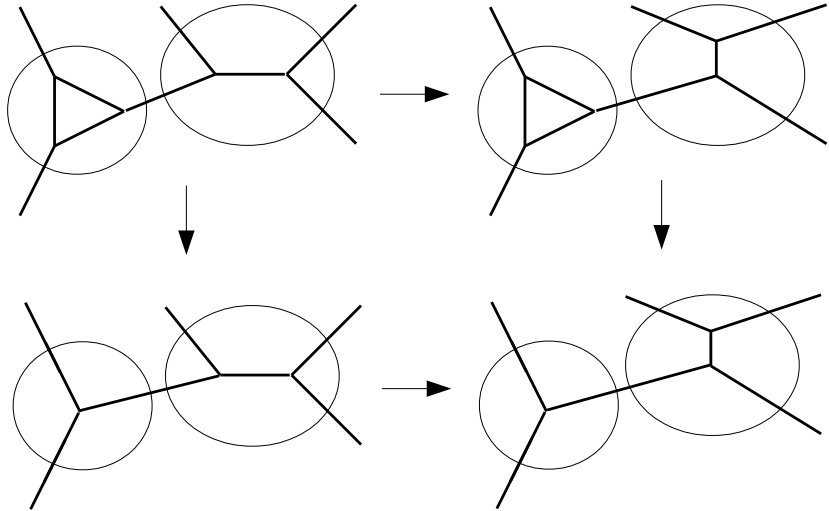}
\caption{ \footnotesize An example of ``two-fingered'' time.  In the upper left corner is a network with two different IER's, each of which can be evolved to a different FER.  (To keep the diagram simple, only one possible FER has been shown for each IER to evolve to.)  There are two different, but equivalent, ways to evolve from the upper left to the lower right.  Each path represents a different way of ``slicing'' the same spacetime history.  In a generally covariant theory, there is no notion of absolute simultaneity, so each path is equally good.  (The arrows in this diagram should not be confused with the quantum indeterminacy arrows shown in Fig. \ref{replace}, in which the arrows represent mutually exclusive histories, each with its own quantum amplitude.)
}\label{threads}
\end{figure}

In quantum gravity, the analogy to threading is that at any given time, there are multiple replacement operations which can act on the spacetime independently (Fig. \ref{threads}).  In such cases, it makes no difference which transition happens ``first''; one can perform them in either order.

\begin{figure}[ht]
\centering
\includegraphics[width=.45\textwidth]{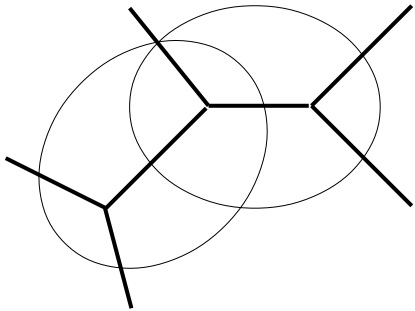}
\caption{ \footnotesize Suppose there are two different overlapping IER's.  This is a disaster because whichever one evolves first will eliminate or change the other IER, causing the histories to diverge permanently.  What determines which IER will succeed in evolving forwards first?  In a generally covariant theory, there is no absolute clock which can mediate between different physical processes by giving them relative speeds.  If the IER is chosen arbitrarily, then deterministic evolution of the wavefunction is lost.  If the IER is chosen on the basis of features of the network outside the intersection of the two IER's, then causality is lost.  (If the conflict could be resolved solely based on information inside the intersection, causality would be saved.  But then it would be better just to say that only one of the regions contains the right data to be an IER, and in fact there are no overlapping IER's.)
}\label{race}
\end{figure}

By contrast, a race condition occurs when two possible replacement rules act on overlapping regions in space  (Fig. \ref{race}).  In such situations, it makes a difference which operation happens ``first''.  But in quantum gravity there is no notion of absolute time to decide which of the events does happen first.  Since time itself arises as a result of the interactions of the discrete units, there is no way to resolve the question of which event happens first, and the quantum gravity model is inconsistent with causality.\footnote{Why not simply sum over all possible orderings of the overlapping regions?  This would seem to reduce the evolution ambiguity to the usual indeterminism of quantum mechanics.  However, I believe that this approach would probably have to violate either causality or unitarity.  Since the probabilities of the different possible choices have to add to 1, it is necessary to somehow assign a probability weight to each evolution region.  For example, if there are $N$ different overlapping IER's and each is equally likely, then each must be assigned a probability of $1/N$ to occur.  But this introduces an element of acausality since the evolution inside of an IER now depends on $N$, which depends on information outside of the IER.  Hence there would no longer be well-defined causal domains of dependence.}  (For a discussion of this issue in the context of a classical model of networks evolving by means of replacement rules, see Wolfram \cite{wolfram}.)

An analogous problem arises in the context of canonical quantum gravity, if the constraint algebra fails to close.  This would result in the Hamiltonian constraint eliminating too many degrees of freedom.  The basic problem is that two different ways of evolving a spatial slice forwards in time need to give the same answer, i.e. their commutator must vanish up to a gauge-equivalence.  This is a highly nontrivial problem in loop quantum gravity, whose solution is still unclear \cite{NPZ05}.  According to Perez \cite{perez05}, all known formulations of the Hamiltonian constraint in which the constraint algebra closes are ``ultralocal'' in the sense of Smolin \cite{smolin96}, meaning that information does not propagate from vertex to vertex.

This problem can be resolved in the case of a ``perfect action'' for which the transition amplitudes are the same (up to gauge symmetry) regardless of which replacement rule acts first \cite{perfect}.  This is the situation for three dimensional loop quantum gravity, which has no local degrees of freedom, and is closely related to a topological Chern-Simons field theory \cite{CS}.  In this case, the spin-foam amplitudes are independent of the choice of triangulation (which is therefore regarded as a gauge symmetry) \cite{3D}.  However, it is unclear whether this approach can be extended to theories with local degrees of freedom, such as general relativity in $D \ge 4$, since it would require the evolution rules to satisfy a complicated set of consistency relationships.

Another proposed solution is the ``master constraint program'' \cite{master} in which the infinitely many constraints of GR are replaced with a single master constraint, which is an integral of the squares of the lapse and shift constraints.  This ensures that the constraints close, but at the price of obscuring the property of foliation independence.  Although classically the master constraint formalism is equivalent to general relativity, quantum mechanically there may be anomalies and so it is desirable to have a formalism in which the foliation independence is manifest.

The DUCT models presented here will solve this problem in a different way.  The dynamical evolution rules will be required to have non-overlapping transitions only.\footnote{One could also construct consistent models with overlapping IER's if each process was ``read-only'' in the intersection.  In other words, if each process only used the intersection of the IER's by measuring a set of commuting quantities, it would not matter which of the two processes occurred first.  The DUCT's constructed here will not use this loophole.}  Hence the ambiguity never arises, because the evolution rules at one moment of time all commute with each other trivially.\footnote{In continuum general relativity, the constraint algebras close nontrivially, since the commutator of the Hamiltonian constraint includes the spatial diffeomorphism constraint.  If any DUCT has semiclassical general relativity as a continuum limit, it would of course be necessary to somehow recover the same algebra of constraints as in general relativity.  While I hope that the DUCT does have a good continuum limit, in this article I am focusing instead on the problem of defining consistent dynamics.}  This allows the time evolution rules of the DUCT to be unitary, causal, and many-fingered.  As a result one can explicitly prove that the space of states is gauge-degenerate under lapse diffeomorphisms (section \ref{Degen}), which is an unsolved problem in the spin foam approach \cite{LFprivate}.

In multithreaded programming, the problem of race conditions is resolved by carefully defining the rules for what each thread can do, in such a way that no conflicts can occur.  For example, a thread might place a ``lock'' on certain bits of information, that only it can access, until it has finished with them.

In quantum gravity, one can accomplish this by adding additional rules so as to specify which vertices are allowed to interact next.  In such a model, the laws of physics would designate certain kinds of network regions as IER's, and other kinds as FER's.  The basic consistency rule is that no two IER's can overlap lest future time evolution be indeterminate.  Nor can two FER's overlap, lest past evolution be indeterminate.  There is, however, no problem if an IER and a FER overlap (in fact they had better sometimes overlap, if information is to propagate from place to place).

\begin{figure}[ht]
\centering
\includegraphics[width=.5\textwidth]{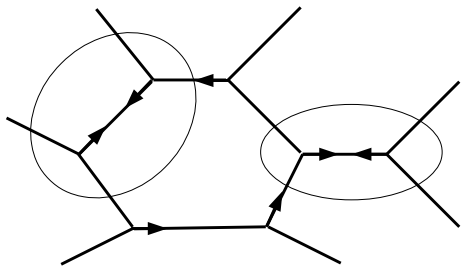}
\caption{ \footnotesize Race conditions can be avoided by adding additional structure to the network.  The simplest way to do this is to equip each vertex with an arrow pointing along one edge.  If two vertices point at each other, they compose an IER.  This rule prevents IER's from overlapping.  An additional kind of arrow (not shown) is needed to define nonoverlapping FER's.}\label{happycouple}
\end{figure}

It is difficult to find reasonable rules for writing down IER's and FER's for spin-network states.  For this reason, I will consider a networks labelled by a different kind of structure (this structure could be either instead of, or in addition to, the spin labels):  Let each vertex be labelled with a ``future arrow'', which points in the direction of one of its edges.  An IER could then consist of a pair of adjacent vertices whose arrows point towards each other (Fig. \ref{happycouple}).
This guarantees that the IER's can never overlap with one another.  The framework can be extended by allowing certain vertices to ``point to themselves'', i.e. to evolve on their own as a one-vertex IER.

Similarly, each vertex must also be labelled with a ``past arrow'' degree of freedom which determines the FER's.
\pagebreak[4]

\begin{figure}[ht]
\centering
\includegraphics[width=.5\textwidth]{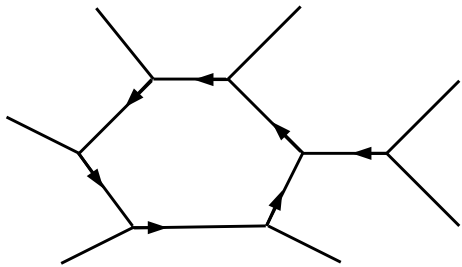}
\caption{ \footnotesize A bunch of vertices which are deadlocked according to the pair rule.  None of the vertices of this hexagon can evolve towards the future, because each vertex has to wait for the next one before it can do anything.  The additional vertex shown pointing into the hexagon is also deadlocked.  The solution is to make the hexagon (or any other set of vertices pointing in a circle) also count as an IER.}\label{deadlock}
\end{figure}

Unfortunately, this rule for defining IER's and FER's immediately leads to a new problem.  What if a bunch of vertices point to each other in a loop (Fig. \ref{deadlock}), so that no pairs are possible?  

If the arrows point in random directions on a network, this sort of situation should arise frequently.  In mutithreaded programming this is known as deadlock, and occurs when some set of threads are waiting for each other before they can continue.  As a result, the program grinds to a halt.  In computer programming this is bad because the program is not supposed to quit before it outputs the answer to the problem.

In quantum gravity, on the other hand, deadlock would mean that time comes to an end, after which no future evolution is possible.  Under the right circumstances, this \emph{could} be an accurate description a final singularity or a Cauchy horizon, either of which would prevent further evolution.  As long as the deadlocked region is shielded from the outside by a horizon (cosmic censorship), time evolution could continue to occur in other regions of the universe.  But apart from strongly gravitating regions, it is important that time should not be able to come to an end (except perhaps with very small probability).  Otherwise why are we still here?

Can an evolution rule be found which is not so conducive to deadlock?  One simple solution can be shown to remove all deadlock from the previous example.   Whenever the future arrows point in a loop as in Fig. \ref{deadlock}, just take that set of vertices composing the loop to be an IER (and similarly for the past arrow and FER's).

The disadvantage of this approach is that the domains of dependence for causality can be arbitrarily large; evolution is not confined to nearest-neighbor interactions.\footnote{I have not been able to find any rule without overlapping IER's or deadlocks without permitting IER's of arbitrary size, but I have not been able to prove that it is impossible, either.} Thus although causality is preserved, the theory is only quasi-local.  The advantage is that the causal domains never overlap, and for any finite closed network there is at least one such region.

Thus in the attempt to construct a theory of quantum gravity with good causality, one is naturally led to a picture in which spacetime is filled with closed string-like causal entities, along which information can instantly propagate.  Nevertheless, there is still many-fingered time, since time has one ``finger'' for every causal string in existence at a given time.  If these strings have a characteristic size on the order of the Planck length, then one would presumably recover a more ``point-like'' causal structure at much larger distances.

\subsection{Outline of this Article}

The plan of this article is as follows: Section \ref{kin} will more precisely define the kinematics of a Discrete Unitary Causal Theory (DUCT).  Section \ref{dyn} will define the unitary dynamics of the theory using a sum over histories approach, paying careful attention to the action of spacetime diffeomorphims on the Feynman rules of the theory.  

Section \ref{IP} will discuss the properties of this inner product.  In order to obtain a physical Hilbert space $\mathcal{H}_\mathrm{phys}$ for some region in the universe, one needs this inner product to be a) hermitian, b) highly degenerate (so as to identify the same ``state'' at different ``times'' as gauge-equivalent) c) positive, and d) finite.  The first two conditions will be shown to hold for all DUCT models.  Positivity will be shown for any region of the universe which had a first moment of time.  No general proof of finiteness will be given; in my view this is likely to depend on the specific properties of the model.

Section \ref{exam} will give some two examples of DUCT's, one for bivalent networks and one for trivalent networks.  Section \ref{pros} will discuss some open questions and prospects for further research. 

Appendix \ref{bound} will place a bound on the divergence in the inner product, showing that it cannot diverge faster than exponentially with respect to the number of transitions in the history.  Appendix \ref{cycle} will discuss quantum interference effects for spacetimes whose histories are cyclical (i.e. periodic) in time.  It will be argued that unless there is a nonzero probability to break out of the cycle, generically there are no dynamical states corresponding to such cyclic histories.

\section{Kinematics}\label{kin}

One can define a ``basis state'' $A,\,B,\,C...$ of a time slice to consist of a network (with arbitrary valency) consisting of a finite set $V$ of vertices, and a set $E$ of symmetric edges, each of which is connects two vertices (or connects a vertex to itself), such that each vertex is equipped with a ``future arrow'' $f(V)$ and a ``past arrow'' $p(V)$, each of which points either (a) along an edge of $V$ towards an adjacent vertex or (b) to the vertex itself.  In this article, I will restrict attention to the set of all $N$-valent networks where each vertex is connected to exactly $N$ edges.  Since a DUCT cannot have processes which break up a connected network into two disjoint spacetimes, one can choose to restrict attention to connected networks.\footnote{This framework can be generalized further in a number of ways, such as by adding colors or orientations to the vertices and edges, imposing additional constraints concerning which combinations of edges can meet at each point, adding gauge fields, giving vertices and edges fermionic statistics instead of bosonic statistics, etc.  One could even generalize the underlying finite topology from a network to something more complicated, like an $n$-complex.  The choice of structure provides the kinematics of the DUCT.  Most of what I say below holds regardless of these choices, but for simplicity I will not consider such modifications here.

There is also no particular reason why the degrees of freedom in the future and past arrows need to commute with each other.  If they do not commute, the future-arrow basis states will differ from the past-arrow basis states by means of a unitary transformation applied to each vertex.  However, the restriction to the commuting case here is actually without loss of generality.  That is because the failure to commute can be incorporated at the level of the dynamics, as will be shown in section \ref{noncom}.}

Define $\mathrm{Iso}(A \to B)$ to be the set of all graph isomorphisms between the network $A$ and the network $B$ (i.e. all finite permutations of the network vertices and edges which  preserve all the structure).  For all $A$ and $B$, if $\mathrm{Iso}(A \to B)$ is nonempty, the two basis states should be regarded as being the same basis state (i.e. $A = B$).\footnote{This assumes bosonic statistics.  If there are fermionic elements, then appropriate minus signs must be inserted.  In particular, if there are any diffeormophisms which relate a network to itself up to an odd permutation of fermionic components of the network, that network would be forbidden by the Pauli exclusion principle and should not be counted as a basis state.}  This equivalence under graph isomorphisms is a discrete analogue of spatial diffeomorphism invariance.  Let the automorphism group $\mathrm{Aut}(A) = \mathrm{Iso}(A \to A)$ consist of the subgroup of diffeomorphisms which send $A$ and all its equipment to itself.

Let a ``kinematic state'' be defined as a complex superposition of these diffeomorphism-invariant basis states.\footnote{These correspond to the ``diffeo\-morphism-invariant'' states of loop quantum gravity.  The term is somewhat of a misnomer, since only spatial diffeomorphisms have been imposed so far.}  There is a natural inner product on these kinematic states:
\begin{equation}\label{KIP}
\langle B | A \rangle_{\mathrm{K}} = |\mathrm{Iso}(B \to A)| = \delta_{AB}|\mathrm{Aut}(A)|.
\end{equation}
The vector space of kinematic states which have finite inner product is therefore a Hilbert Space.

Eq. (\ref{KIP}) may be viewed as a primitive form of a sum over histories where the only ``histories'' permitted are spatial diffeomorphisms.  The factor of $|\mathrm{Aut}(A)|$ counts the number of histories relating two slices.  To motivate this factor, note that the kinematic inner product invariant would not be invariant under local unitary changes of basis acting on the data at each vertex.\footnote{As a particularly simple example, consider a network consisting of just two vertices connected by a single edge, which possesses a $\mathcal{Z}_2$ symmetry under permutation of the two vertices.  Let each vertex be labelled by one of two states, either $0$ or $1$, which are regarded as basis states in a two-dimensional Hilbert Space having the inner product $\langle a | b \rangle = \delta_{ab}$ where $a,b = 0,1$.  The three possible diffeomorphism-invariant states of the whole system are $|00\rangle$, $|01\rangle = |10\rangle$, and $|11\rangle$.  Then one can act on each vertex with a unitary replacement rule such as 
$|a\rangle = (|0\rangle + |1\rangle)/\sqrt{2}$, 
$|b\rangle = (|0\rangle - |1\rangle)/\sqrt{2}$.  
One finds that the inner product is not preserved unless the norm-squared of the symmetrical states such as $|00\rangle$ is doubled relative to the norm-squared of the asymmetrical states such as $|01\rangle$.  Then one has e.g:
\begin{equation}
2 = \langle aa | aa \rangle = \frac{1}{4} \langle 00 | 00 \rangle + \langle 01 | 01 \rangle + \frac{1}{4} \langle 11 | 11 \rangle = \frac{1}{2} + 1 + \frac{1}{2}.
\end{equation}
Without the factors of $2$ one would have $1 = 1/4 + 1 + 1/4$ which is not consistent.}

One can also define basis states and the kinematic inner product for network regions with a boundary $Q$ containing some number of external edges.  The procedure is exactly the same except that the automorphism group should be restricted to symmetries which do not affect the boundary.  It will be particularly important to define inner products on the space of IER's, and also the space of FER's, since these inner products are needed to define unitarity for time evolution from the space of IER's to the space of FER's. 

Define a IER ``basis state'' $i,\,j,\,k...$ to consist of any $n$ vertices which can be mapped onto the natural numbers from $1...n$ such that for all $m$, the $m$th vertex has its future arrow to the $(m+1)$th vertex modulo n.  A FER ``basis state'' $a,\,b,\,c...$ is defined similarly but in terms of the past arrows.  The IER's and FER's can be grouped into ``boundary classes'' based on the total number of edges connecting the IER or FER to the rest of the network.  The complex superpositions of these networks with boundary can be made into IER and FER Hilbert spaces in the same way as the total network states, by applying the appropriate analogues of Eq. (\ref{KIP}):
\begin{eqnarray}\label{KFP}
\langle j | i \rangle_{\mathrm{K}} = \delta_{ij}|\mathrm{Aut}(i)| = K_{j'i}, \\
\langle b | a \rangle_{\mathrm{K}} = \delta_{ab}|\mathrm{Aut}(a)| = K_{b'a},
\end{eqnarray}
where the automorphism group excludes symmetries which affect the boundary.  The last equality of each line introduces a tensor-like notation in which unprimed lower indices represent kets and the primed lower indices represent bras.  This notation will be useful below because of the need to raise and lower indices using the inner product $K$.  The inverse kinematic inner product will be written as $K^{j'i}$ or $K^{b'a}$.

In summary, kinematic Hilbert spaces can be constructed for a) network states without boundary, b) network states with a boundary $Q$, which includes as special cases c) IER's with boundary $Q$ and d) FER's with boundary $Q$.  Compatibility between (a), (c), and (d) requires that the factors of $|\mathrm{Aut}(A)|$, $|\mathrm{Aut}(i)|$, or $|\mathrm{Aut}(a)|$ be included in these inner products; otherwise the proof of gauge degeneracy found in section \ref{Degen} will not work.

\section{Dynamics}\label{dyn}

The next step is to introduce the invertible unitary transformations $U$ which will be used to evolve one network into the other by substituting IER's for FER's, and vice versa (section \ref{transit}).  These unitary transformations can then be used to construct a sum-over-histories formalism.  By choosing the right Feynman rules for the theory, the inner product between any two states can be constructed (section \ref{rules}).  Even if the sum over histories is not absolutely convergent, one may still obtain good results if Abel resummation is possible (section \ref{resum}).

In loop quantum gravity, the sum over histories is normally derived from the Hamiltonian constraint by the procedure known as ``group averaging'' or the ``rigging map'' \cite{Gavg}.  This strategy will not be used in what follows.\footnote{For one thing, the discrete lapse diffeomorphisms of the DUCT do not form a group.}  Instead, the Feynman rules will be postulated to satisfy certain causality constraints and symmetry equivalences.  Like group averaging, these rules will have the consequence of restricting the space of kinematic states to a much smaller dynamical Hilbert Space.  (Because I am working with the dual space of states, this reduction is given not by projecting onto a constraint, but by modding out pure gauge vectors, but the principle is the same.)  Thus, although the fundamental building blocks of the DUCT are unitary transformations $U$, the sum over histories involves the reduction of the state space to a quotient space using a highly degenerate inner product.\footnote{Given that fundamentally there is no notion of absolute time in a background independent theory, one might wonder why the $U$ matrices should be unitary.  One reader writes that ``Unitarity is usually required to preserve the inner product (in systems with an external time).  In systems without such a time it is unclear why one wishes for unitarity.''  However, if one replaces $U$ with a nonunitary matrix, the proof of gauge degeneracy found in section \ref{Degen} is no longer valid.  Hence the unitarity of $U$ is in fact important for consistency of the dynamics.  From a conceptual point of view, unitarity expresses the idea that any kinematic state may be evolved either to the past or to the future in a way that conserves probability.}

\subsection{Unitary Transitions}\label{transit}

The dynamics are fully specified by the following choice: for each boundary class $Q$ as defined in the previous section, select an invertible unitary matrix $U$ which is symmetric with respect to any symmetries of $Q$.  The domain of $U$ is the vector space of all complex superpositions of IER basis states whose boundary is $Q$, and the target of $U$ is the analogous vector space of FER basis states whose boundary is $Q$.

If one defines the adjoint of $U$ by raising and lowering with respect to the kinematic inner product, and then complex conjugating (which switches ket and bra indices):
\begin{equation}
U^{\dagger i}_{\phantom{\dagger} a} = \sum_b \sum_j (U^b_j K^{i'j} K_{a'b})^*,
\end{equation}
then the condition that $U$ is unitary and invertible means that
\begin{equation}
\sum_a U^{\dagger i}_{\phantom{\dagger} a} U^a_j = \delta^i_j \, ; \quad 
\sum_i U^a_i U^{\dagger i}_{\phantom{\dagger} b} = \delta^a_b.
\end{equation}
Because the kinematic inner product is nontrivial, the matrix $U$ does not look unitary when written in terms of the matrix elements relating the IER and FER basis states.  However, it can be related to an orthonormal basis as follows:
\begin{equation}\label{I}
U^a_i = \sqrt{ \frac{|\mathrm{Aut}(i)|}{|\mathrm{Aut}(a)|} }\,U_{\hat{a} \hat{\imath}}.
\end{equation}
where for hatted indices there is no difference between lower and upper indices.

The matrix $U$ can be used to evolve a IER forward in time to a FER, and its inverse 
$U^{-1}\!= U^\dagger$ can be used to evolve a FER backwards in time to a IER.  Since time evolution is a gauge symmetry, $U$ is to be regarded as expressing a physical equivalence between the different initial and final evolution regions.  The invariance of the physical inner product under this identifiction will be shown explicitly in section \ref{Degen}.

It will also be convenient to define a transition matrix $T$ defined by lowering $U$ with respect to the kinematic inner product:
\begin{equation}\label{T}
T_{a'i} = \sum_b K_{a'b} U^b_i = \sqrt{ |\mathrm{Aut}(a)| \cdot |\mathrm{Aut}(i)| }\,U_{\hat{a} \hat{\imath}}.
\end{equation}
This matrix has the nice property that it is obtained from $U$ in a way that is symmetric under time reversal (although $U$ itself might not have time reversal symmetry).

\subsection{Feynman Rules for the Sum over Histories}\label{rules}

The dynamics should have the effect of radically lowering the number of degrees of freedom of the system, since any two superpositions of slices which are related by time evolution are actually gauge-equivalent to one another.  To implement this reduction of degrees of freedom, the dynamical inner product will be defined as a sum over all the histories relating two different basis states $A$ and $B$:
\begin{equation}\label{DIP}
\langle B | A \rangle = \sum_{H(A, B)} \mathcal{A}(h),
\end{equation}
where $H(A, B)$ is the class of histories going from $A$ to $B$ and $\mathcal{A}(h)$ is the amplitude associated with a particular history, defined below.  Subject to some additional restrictions to be described momentarily, a ``history'' $h$ in the class $H(A, B)$ is a sequence of basis states $h_n$, $0 \le n \le t$, starting with $h_0 = A$ and ending with $h_t = B$, such that each basis state $h_n$ is obtained from the previous basis state $h_{n-1}$ by substituting an IER for a FER of the same class, or vice versa.\footnote{In general, one must allow histories which contain both forwards and backwards evolution.  That is because each $\mathrm{IER} \to \mathrm{FER}$ unitary transition represents a gauge identification between two different kinematic states, and identity is a symmetric relationship.  If $A = B$ it must equally well be true that $B = A$, regardless of the chronological order of the slices $A$ and $B$ in the spacetime.  The compatibility of the inner product with this gauge identification will be proven in section \ref{Degen}.

Contrary to Ref. \cite{teitelboim83}, the admission of backwards-evolving histories is in no way opposed to the principle of causality.  Despite the use of the suggestive names ``initial slice'' and ``final slice'',  there is no reason why one time slice cannot lie partly or entirely to the past of some other time slice.  After all, causality in general relativity is perfectly consistent with the existence of slicings which cross each other in arbitrary ways.  Indeed, as shown in Ref. \cite{teitelboim83}, the admission of both forwards- and backwards-evolving histories is \emph{required} for the path integral to be generally covariant.}  Since the theory is background free one sums over all possible (nonnegative integer) values of $t$ as well as all possible histories with a given $t$.

Every history defines a causal structure between its vertices and transitions as follows.  Whenever a IER transitions to a FER (or vice versa), all of the vertices in the IER may be defined as being to the future of the transition, while all the vertices in the FER may be defined as being to the past of the transition.  Then there is a partial ordering which is generated by all these causal relations plus those which follow from the transitivity of causality (i.e. if $x > y$ and $y > z$, $x > z$).

In writing down the exact Feynman rules, it is necessary to have precise rules for which histories are distinct and which are not, so as to avoid double counting gauge equivalent processes.  The following rules give the necessary information.  (The justification for these rules is implicit in the next section's proofs concerning the nice properties of the inner product---if these rules were changed the proofs would not work.)  \textit{The notation $A \!\smile\! B$ will be used to describe any region formed by gluing two regions $A$ and $B$ together along part (or all) of their boundaries.}

Accordingly, the definition of a history is subject to the following additional restrictions and identifications:
\begin{enumerate}
\item \textbf{No Backtracking.} It is not permissible for a history to backtrack, by transitioning from a IER to a FER and then from that same FER back to an IER (or vice versa), thereby evolving both forwards and backwards in time in the same spatial location.  Such backtracking histories are not counted in Eq. (\ref{DIP}) due to redundancy with non-backtracking histories.  (If such backtracking were allowed, every state with at least one IER or FER would have infinite norm due to backtracking arbitrarily many times along a single transitition.)

A history can contain both forwards and backwards evolution, but only if they occur in spatially separated regions.  That is, the intial and final slices $A$ and $A^\prime$ can straddle one another in the sense that if $A = B \!\smile\! C$ and $A^\prime = B^{\prime} \! \smile\! C^{\prime}$, then $B$ evolves forwards in time to $B^{\prime}$ and $C$ can be evolved backwards in time to $C^{\prime}$.  See Fig. \ref{backtracking}.

\begin{figure}[ht]
\centering
\includegraphics[width=.8\textwidth]{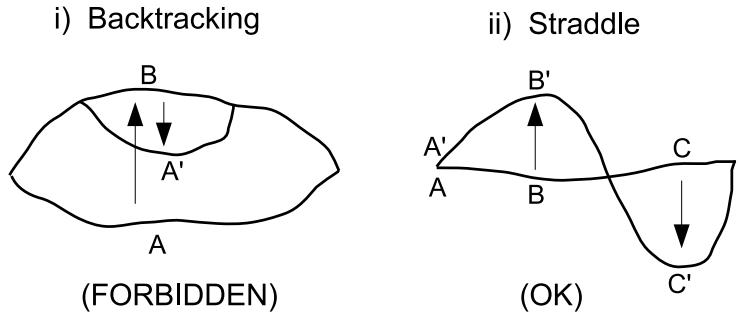}
\caption{ \footnotesize Spacetime diagrams of histories going from a slice $A$ to a slice $A^\prime$ by means of processes that involve both forwards and backwards time evolution.  Part (i) shows a history which evolves forwards to an intermediate slice $B$, and then backwards again to $A^\prime$.  This is backtracking, which is excluded from the sum over histories due to redundancy (since the relevant amplitude is already contained in the nonbacktracking $A \to A^\prime$ histories).  Part (ii) shows a straddling history in which one region of space is evolved forwards in time, while another region of space is evolved backwards in time.  These histories must be included in the sum, in order for the inner product to be diffeomorphism-invariant.}\label{backtracking}
\end{figure}

\item \textbf{Lapse Diffeomorphisms.} Two histories are to be identified if they differ only by a change in the order in which two transitions in the sequence are applied, so long as the two transitions are spacelike separated.  Histories which can be identified by any chain of such permutations are to be considered the same history and counted only once in Eq. (\ref{DIP}).

\item \textbf{Shift Diffeomorphisms.} Any two histories which differ only by permutations of vertices and edges in intermediate basis states are to be regarded as equivalent, so long as the permutations preserve the causal structure and the network relationships.  However, two histories are to be regarded as distinct if they differ by a permutation of the initial or final basis states.  (Otherwise there would be an extra factor of $(|\mathrm{Aut}(A)| \cdot |\mathrm{Aut}(B)|)^{-1}$ in $\langle B | A \rangle$ which prevents histories from composing properly, and the proof of gauge-invariance in section \ref{Degen} would not work.)

\item \textbf{Initial and Final Symmetries.}  To count the number of distinct histories arising from the fact that automorphisms applied to the initial and final slices must be considered distinct histories, one may label the network vertices or edges of the initial state $A$  of the history $h$ with labels from $1...n$ in an arbitrary manner, where $n$ is the total number of vertices and edges in $A$.  Similarly, arbitrarily label the vertices of the final state $B$ with numbers from $1...m$ where $m$ is the number of vertices and edges in $B$.  Now consider all other ways of labelling the initial and final state of $h$ which differ from the original labelling by applications of $\mathrm{Aut}(A)$ to the initial state and $\mathrm{Aut}(B)$ to the final state.  Count the number of generated histories $h$ which are distinct under spacetime diffeomorphisms, and consider each one a distinct history.  Note that in the case of the trivial history without any transitions, the number of distinct histories is $|\mathrm{Aut}(A)|$ as required to match Eq. (\ref{KIP}).\footnote{If the DUCT were to contain entities with fermionic statistics, these symmetry rules would need to be modified by the appropriate insertions of minus signs for fermion exchange and fermion loops.}
\end{enumerate}
The amplitude $\mathcal{A}(h)$ associated with each history is given by the following formula:
\begin{equation}\label{amp}
\mathcal{A}(h) = \frac{1}{|\mathrm{Aut}(h)|}
\prod_{n = 0}^{t-1} T(h_n \to h_{n+1}).
\end{equation}
where the transition amplitude $T(h_n \to h_{n+1})$ equals $T_{a'i}$ whenever $h_{n+1}$ is obtained from $h_n$ by evolving a IER in state $i$ forwards in time to a FER in state $a$, and $T^\dagger_{i'a} = (T_{a'i})^{*}$ whenever a FER in state $a$ is evolved backwards in time to a IER in state $i$.  $\mathrm{Aut}(h)$, which is used to determine the symmetry factor in Eq. (\ref{amp}), is the group of automorphisms of the history under lapse and shift diffeomorphisms (as defined above).\footnote{The symmetry factors $1/|\mathrm{Aut}(h)|$ (in Eq. (\ref{amp})) and $\sqrt{|\mathrm{Aut}(a)| \cdot |\mathrm{Aut}(i)|}$ (in Eq. (\ref{T})) can be derived by assigning a continuous dummy parameter $0 < \eta < 1$ to each vertex and edge of the basis states.  Each basis state (including those used to define IER's and FER's) can then be defined by a uniform wavefunction over all possible values of $\eta$.  Because the $\eta$ parameter is continuous, it is infinitely unlikely that any two network elements will have the exact same value of $\eta$, therefore the $\eta$ parameter breaks all symmetries of both networks and histories.  However, after integrating over all possible $\eta$ values, one recovers the Feynman rules stated above, including the symmetry factors.}

By putting Eq. (\ref{amp}) together with Eq. (\ref{DIP}) one finds that
\begin{equation}\label{DIP2}
\langle B | A \rangle = \sum_{H(A, B)} 
\frac{1}{|\mathrm{Aut}(h)|}
\prod_{n = 0}^{t-1} T(h_n \to h_{n+1}).
\end{equation}

The same set of rules can also be adapted to derive an inner product for spatially bounded regions, holding the boundary $Q$ fixed.  In this case, histories should be regarded as distinct if they differ by a symmetry which affects $Q$.  States may then be then classified by means of irreducible representations of $\mathrm{Aut}(Q)$.  By virtue of the symmetry, states in different irreps of $\mathrm{Aut}(Q)$ must have zero inner product with each other.  

Unfortunately, it is not usually possible to recover the Hilbert space of the whole spacetime by tensoring the Hilbert space of the interior of $Q$ with the Hilbert space of the exterior of $Q$.  That is because the tensor product does not take into account diffeomorphisms which affect the boundary $Q$.

\subsection{What if the Sum does not Converge?}\label{resum}

Because of the oscillatory nature of quantum mechanics, it might happen that, for a particular choice of parameters, this sum over histories is not absolutely convergent.  In such cases it may be necessary to use a resummation technique to obtain finite answers.  

Fortunately, if there is a divergence, it is not very severe.  Assuming that there is a maximum number $M$ of states allowed to mix in any $\mathrm{IER} \to \mathrm{FER}$ transition---which is true for the particular DUCT's constructed in section \ref{exam}---the sum of the absolute amplitudes of each history can diverge no faster than exponentially with respect to the maximum number of transitions $t$ allowed in the history (see appendix A).  This is good, since it suggests that the total amplitude \emph{might} converge for favorable values of the parameters of the DUCT.  (By contrast, if the number of possible histories had grown faster than exponentially with $t$, the amplitude would always diverge, assuming that the amplitude of a typical history falls off exponentially with $t$.)  This nice property is probably due to the existence of a causal structure, since similar results are reported for causal dynamical triangulations \cite{AL98}.

If the sum over histories does not converge, it may be necessary to use Abel resummation to extract a finite answer.  To do this, introduce a ``coupling constant'' $0 \ge c \ge 1$ and consider a modification to Eq. (\ref{DIP2}) coming from multiplying by $c$ once per transition:
\begin{equation}\label{abel}
\langle B | A \rangle = \sum_{H(A, B)}
\frac{c^t}{|\mathrm{Aut}(h)|}
\prod_{n = 0}^{t-1} T(h_n \to h_{n+1}),
\end{equation}
which is equal to Eq. (\ref{DIP2}) if the latter is well-defined.  Since the sum over histories diverges at most exponentially, this function must converge for sufficiently small values of $c$.  Since it is defined as a power series, it is by definition analytic within its radius of convergence.  Assuming this analytic function can be continued to $c = 1$ along the real axis without encountering an obstruction, this provides an alternative definition of the inner product.\footnote{See appendix \ref{quant} for a simple case in which Abel resummation can be used to interpret a sum which is not absolutely convergent.}

All the nice properties of Eq. (\ref{DIP2}) which will be shown in section \ref{IP} also apply to the Abel resummed inner product, given one additional technical assumption stated in section \ref{Degen}.\footnote{Basically, the proof of degeneracy in section \ref{Degen} requires dividing a set of histories into two classes based on whether time evolution goes forwards or backwards through a particular IER/FER transition.  The additional assumption is just that each class can be Abel resummed \emph{separately} to get a finite answer.}

\section{Properties of the Inner Product}\label{IP}

The next task is to show that the dynamical inner product of Eq. (\ref{abel}), as defined by the Feynman rules in the previous section, has physically reasonable properties.  In the present section, I will derive the properties of the inner product on the assumption that it is finite when evaluated between any two basis states.\footnote{The inner product can then be used to define which infinite superpositions of basis states are normalizable.}  In other words, it will be assumed that this inner product is either absolutely convergent, or else can be resummed in a way that is compatible with the proof of these properties.  The validity of this assumption may depend on the precise form of the dynamical laws.

First the inner product will be shown to be hermitian and to have a large amount of degeneracy.  This degeneracy represents the gauge-invariance of the theory under spacetime diffeomorphisms, which are described by the action of the $U$ matrix acting on any IER of a basis state.  States which are gauge-equivalent to the zero vector can then be modded out to construct the dynamical state space, which is invariant under all spacetime diffeomorphisms.

Then it will be shown that, at least for states of the universe which began at a finite time in the past, the inner product on the dynamical state space is positive-definite, making it a Hilbert space.  This gives the theory a probability interpretation via the Born rule.

\subsection{Hermitianness}

First note that the inner product (\ref{DIP2}) is hermitian:
\begin{equation}\label{herm}
\langle A | B \rangle = \langle B | A \rangle^*,
\end{equation}
since for every history going from $A$ to $B$, there is a time-reversed history going from $B$ to $A$ obtained by reversing all the transitions and taking the complex conjugate of all the transition amplitudes.  This also applies to the Abel resummed inner product (\ref{abel}).

\subsection{Gauge Degeneracy}\label{Degen}

Secondly, the inner product (\ref{DIP2}) has a large amount of degeneracy, because any time a network takes the form $C \!\smile\! i$ where $i$ is a IER, $C$ is the complement to the IER, and $B$ is any other basis state, the following inner product vanishes:
\begin{equation}\label{ketdeg}
\langle B |
\left( | C \!\smile\! i \rangle - \sum_a U^a_i | C \!\smile\! a \rangle \right) = 0.
\end{equation}
This degeneracy is very important because it means that the number of dynamical states is much lower than the number of kinematic states.  Without this gauge symmetry relating seemingly different states, there could be no nontrivial dynamics in the theory.

One can also write Eq. (\ref{ketdeg} in the FER basis by inverting $U$:
\begin{equation}
\langle B |
\left( | C \!\smile\! a \rangle - \sum_i (U^{-1})^i_a | C \!\smile\! i \rangle \right) = 0,
\end{equation}
or in terms of bra states by complex conjugating $U$:
\begin{equation}\label{deg}
\left( \langle C \!\smile\! i | 
- \sum_{a'} U^{*a'}_{\phantom{*}i'} \langle C \!\smile\! a| \right) |B \rangle = 0.
\end{equation}
Since these degeneracy relationships are all equivalent, it will suffice to prove Eq. (\ref{deg}).

The proof of Eq. (\ref{deg}) which follows works because: 1) $U$ is unitary, and 2) the factors of $|\mathrm{Aut}(i)|$ and $|\mathrm{Aut}(a)|$ appear in the right places so long as the sum over histories is defined using the rules of section \ref{dyn}.  This justifies the choices made in that section.  I will start by assuming absolute convergence of the inner product defined by Eq. (\ref{DIP2}), and then discuss the generalization to the Abel resummed inner product Eq. (\ref{abel}).

Proof of Eq. (\ref{deg}): By linearity of the inner product rewrite Eq. (\ref{deg}) as
\begin{equation}\label{E}
\langle C \!\smile\! i | B \rangle = \sum_{a'} U^{*a'}_{\phantom{*}i'} \langle C \!\smile\! a | B \rangle.
\end{equation}
Contributions to the l.h.s. of Eq. (\ref{E}) consist of histories going from $B$ to $C \!\smile\! i$.  Such contributions divide into two kinds: 1) the IER $i$ is obtained by evolving backwards in time from some FER state $a$, or 2) those in which it is not.

Similarly, contributions to the r.h.s. of Eq. (\ref{E}) consist of histories going from $B$ to $C \!\smile\! a$ for some choice of $a$.  These contributions divide into three kinds: 1) those in which the FER $a$ is not obtained by evolving forwards from any IER state, 2) those in which the FER $a$ is obtained by evolving forwards from the IER state $i$, and 3) those in which the FER $a$ is obtained by evolving forwards from some other IER state $j \ne i$.

The strategy of the proof is to show that for every contribution to the l.h.s., there is an equal contribution to the r.h.s.  Specifically, each contribution of the form (1) to the l.h.s. is balanced by an equal contribution of the form (1) to the r.h.s., contributions of the form (2) to the l.h.s. are balanced by equal contributions of the form (2) to the r.h.s., and contributions of the form (3) cancel out.  Let us consider these three cases in detail:

Case \#1: If the IER $i$ is not obtained by backwards evolution, then for every $a$, there exists a modified history $\tilde{h}(a)$ consisting of $h$ plus an $i \to a$ transition to the basis state $C \!\smile\! a$ with amplitude $T_{ai}$.  It may be that $i$ has a nontrivial symmetry group $\mathrm{Aut}(i)$.  For purposes of counting histories from $B$ to $C \!\smile\! i$, each automorphism of $i$ produces a distinct history.  However, when counting histories from $B$ to $C \!\smile\! a$, each  element of $\mathrm{Aut}(i)$ will either result in the identification of two different histories, or else act as an automorphism on a single history.  Either way there is a symmetry factor of $1/|\mathrm{Aut}(i)| = K^{i'i}$ relative to the history $h$.  Putting all of this together, the contribution of the histories $\tilde{h}(a)$ to the r.h.s. of Eq. (\ref{E}) is given by
\begin{equation}\label{case1}
\sum_{a'} U^{*a'}_{\phantom{*}i'} T_{a'i} K^{i'i} \mathcal{A}(h) =
\sum_a U^{\dagger i}_{\phantom{\dagger}a} U^a_i \mathcal{A}(h) = \mathcal{A}(h),
\end{equation}
which exactly matches the contribution of $h$ to the l.h.s.

Case \#2: If the FER $i$ is obtained by backwards evolution from some basis state $C \!\smile\! a$, then in this case one can define a modified history $\tilde{h}$ by removing from the history $h$ the $a \to i$ transition.  Each history $\tilde{h}$ has an amplitude of $\mathcal{A}(h) / T^\dagger_{ia}$, times an extra symmetry factor of $|\mathrm{Aut}(a)| = K_{a'a}$ relative to $h$ due to the fact that the automorphic histories are counted as distinct for purposes of counting histories from $B$ to $C \!\smile\! a$, but not for purposes of counting histories from $B$ to $C \! \smile\! i$, as argued in the preceding case.  Putting all these factors together, the contribution of $\tilde{h}$ to the r.h.s. is
\begin{equation}\label{case2}
\frac{U^{*a'}_{\phantom{*}i'} K_{a'a}}{T^\dagger_{a'i'}} \mathcal{A}(h) 
= \frac{T^\dagger_{a'i'}}{T^\dagger_{a'i'}} \mathcal{A}(h) 
= \mathcal{A}(h),
\end{equation}
again matching the contribution of $h$ to the l.h.s.

Thus every contribution to the l.h.s. is balanced by histories which equally contribute to the r.h.s.  The only remaining issue is whether there exist any other histories contributing to the r.h.s. besides the ones considered above.

Case \#3: The only class of histories going from $B$ to $C \!\smile\! a$ which cannot have an $i \to a$ transition either added or removed from them are those in which $C \!\smile\! a$ is obtained by evolving some other basis state $j \ne i$ forwards in time to reach $a$.  The contribution of this class of histories to the r.h.s. is proportional to 
\begin{equation}
\sum_{a'} U^{*a'}_{\phantom{*}i'} T_{a'j} \propto \sum_a U^{\dagger i}_{\phantom{\dagger} a} U^a_j = \delta^i_j = 0.
\end{equation}
Therefore Eq. (\ref{deg}) is true.  Q.E.D.

\paragraph{Abel resummation.}  In the case where the inner product requires Abel summation, Eqs. (\ref{case1}) and (\ref{case2}) only hold up to a factor of the coupling constant $c$.  This means that there is no degeneracy for $c < 1$.  However, degeneracy may be restored in the physical case $c = 1$.  This happens if the histories contributing to Case \#1 and the histories contributing to Case \#2 each have a finite Abel resummation when considered separately.

\subsection{The Dynamical State Space}\label{DSS}

Armed with the degeneracy shown in the section \ref{Degen}, it is now possible to construct the dynamical (or physical) state space by modding out the kinematic state by all the zero vectors given in Eq. \ref{deg}.  This results in a smaller vector space which inherits an inner product from the old one.  Since the degeneracy can be thought of as encoding the fact that IER's can evolve to FER's, this smaller vector space can be interpreted as the states which are invariant under spacetime diffeomorphisms.  Each dynamical state corresponds to an orbit in the kinematic state space.  The basis states of the kinematic state now form an overcomplete basis of the physical state space.

As stated in the Introduction, by taking the dual of the vector space of states, one can recover the more ordinary picture in which the physical state space is obtained by constraining the kinematic state to lie in the kernel of some constraint operator.

\subsection{Positivity}\label{pos}

The space of dynamical states is not a Hilbert space unless the dynamical inner product defined on the physical states is positive-definite.  If there are other zero vectors other than those identified in Eq. (\ref{deg}), that would not be a big problem because one can just mod out by those additional zero vectors.  But if there are any states which have negative inner product with themselves, the probability interpretation of the theory is in trouble, unless one can find a physical reason to project onto a smaller subspace of states which does have positive probability.

I do not know whether it is possible to prove that the inner product (\ref{DIP2}) is always positive on all dynamical states.  It is however possible to show that it is positive on a certain physically important subspace of states, namely those which have a first instant of time.  Admittedly, since the rules stated in the sections \ref{motive} and \ref{kin} prevent deadlock in both time directions, there are no complete spatial slices which have a first instant of time.  Nevertheless, there still exist spatially bounded regions with a first moment of time.  (There might also exist first moments of time for the whole universe if the evolution rules were modified to be more restrictive, so as to permit deadlock in certain situations.  As stated in the Introduction, the phenomenon of deadlock may be reasonable so long as it occurs in the context of singularities.)

Let us define an ``original'' region of the universe as one which contains no FER's and therefore cannot be evolved into the past.  An ``original kinematic state'' is any complex superposition of original basis states, while an ``original dynamical state'' may be defined as any state which has an original state in its gauge-orbit on the space of kinematic states.  The space of original states is a vector subspace, hence there must exist some projection operator $P$ which projects onto the space of original states.

If the whole universe had a beginning, then it is presumably in an original state (since it is gauge-equivalent by time translation to its beginning, which cannot be evolved further back in time).  This property of originality must be inherited by all of the regions in the universe, since if a region $R$ can be evolved back in time infinitely by a series of unitary evolution steps, then so can any larger region $R^\prime$ which contains $R$.  In such a universe, the only physical states one needs to be concerned with are original states.

Now it will be shown that original states must have a positive dynamical inner product, i.e. for any original state $\Psi \ne 0$,
\begin{equation}
\langle \Psi |\Psi \rangle > 0.
\end{equation}
The dynamical inner product between two states can be calculated using any representative of that state on the kinematic state space.  Let the representative chosen be a complex superposition of original kinematic basis states $O_n$:
\begin{equation}
|\Psi \rangle = \sum_n | O_n \rangle.
\end{equation}
But the inner product between such states is just the kinematic inner product:
\begin{equation}\label{OIP}
\langle O_m | O_n \rangle = \delta_{mn} |\mathrm{Aut}(O_n)|.
\end{equation}
This is because an original kinematic state by definition contains no FER's in it.  But every nontrivial evolution between two regions must involve at least one IER in one of them.  This leaves only the spatial automorphisms, which are counted by the kinematic inner product (\ref{KIP}).  But this inner product is explicitly positive.

I have not found an example of a non-original state with negative norm-squared.  It may be that the dynamical inner product is always positive, even for non-original states.

On the other hand, if there are states with negative norm-squared, the theory would seem not to have a sensible probability interpretation. Yet it might still be possible to make sense out of the theory by restricting attention to original states.  The projector $P$ would then define a Hilbert space of physical states, at least for regions with boundary.  For any region, one would then have to be careful to only act on that region's Hilbert space with operators that commute with $P$.  Since the property of being original is not inherited backwards from a region to the universe as a whole, this would mean that there are operators which are observables for a particular region, but not for the whole spacetime, even when the location of the region can be defined in a diffeomorphism-invariant way.

The prescription would also be non-CPT invariant, which could however be a virtue if it helps to explain the cosmological arrow of time \cite{penrose}.  For example, if the observable universe is inside an original region, then it can contain future horizons, but not past horizons (such as white holes).

\section{Examples}\label{exam}

In this section two examples of DUCT's will be provided for illustrative purposes.  One DUCT will be defined using bivalent networks, and will therefore describe a 1+1 dimensional quantum gravity theory.  The second will be defined using trivalent vertices, suitable perhaps for describing higher dimensional quantum gravity.

Assuming the sum over histories yields a finite nonnegative inner product, these DUCT's serve as examples of consistent microscopic theories of discrete causal spacetimes.  It is not at all clear, however, whether these DUCT's can give rise to approximately classical macroscopic spacetimes.  It also seems difficult for Lorentz invariance to coexist with the discrete spacetime of a DUCT.

In the DUCT models defined below, the past and future arrows do not commute with each other.  This may seem to contradict the definitions in section \ref{kin}.  However, the next section will show that the restriction to commuting arrows is actually without loss of generality.  Hence all the nice properties of the inner product proven in section \ref{IP} also hold for noncommuting DUCT's.

\subsection{Noncommuting Arrows}\label{noncom}

In section \ref{kin} it was assumed that there was a well-defined set of ``basis'' states in which the IER's and FER's were both determinate.  One can generalize this further by permitting the past and future arrows to be noncommuting degrees of freedom.  However, it turns out that every DUCT with noncommuting past and future arrows is actually equivalent to a DUCT in which the arrows commute, justifying the restriction in section \ref{kin}.  This section explains how to define a noncommuting DUCT, and how to construct the equivalent commuting DUCT.

Suppose we consider a theory with $N$-valent vertices.  For each vertex, there is a finite dimensional Hilbert space $\mathcal{H}_\mathrm{local}$ describing the degrees of freedom living at that vertex.  If the edges have no color labels, then $\mathcal{H}_\mathrm{local}$ must be some representation of the permutation group $\mathcal{P}_N$ which permutes the edges.

In order to define the future evolution arrow, one must be able to decompose $\mathcal{H}_\mathrm{local}$ into the direct product of (n + 1) smaller Hilbert spaces:
\begin{equation}
\mathcal{H}_\mathrm{local} = \mathcal{H}_{0}^{\mathrm{fut}} \oplus \sum_{n = 1}^N \mathcal{H}_n^{\mathrm{fut}},
\end{equation}
such that states in $\mathcal{H}_{0}^{\mathrm{fut}}$ have their future arrow pointing to themselves, and states in 
$\mathcal{H}_{0}^{\mathrm{fut}}$ have their future arrow pointing along the $n$th edge coming out from the vertex.  This decomposition must be performed in a way which respects the permutation symmetry of the edges.  One could think of this decomposition as coming from a ``future-arrow operator'' $F$ whose spectrum is an integer from $0...N$.

Similarly, in order to define the past evolution arrow, one must be able to perform a decomposition which is defined the same way, but with respect to the past arrow:
\begin{equation}
\mathcal{H}_\mathrm{local} = \mathcal{H}_{0}^{\mathrm{past}} \oplus \sum_{n = 1}^N \mathcal{H}_n^{\mathrm{past}},
\end{equation}
which can be described by a ``past-arrow operator'' $P$ which also has the spectrum $0...N$.

In general, there is no reason why $P$ and $F$ should commute with each other.  However, restricting them to commute actually leads to no loss of generality.  This is because it is always possible to write a noncommuting model in terms of an equivalent model, in which the failure to commute occurs at the level of the dynamics, instead of the kinematics.

In order to transform a theory in which $F$ and $P$ do not commute into a theory in which they do, simply define a new Hilbert space with twice the dimension as follows:
\begin{equation}
\mathcal{H}^\prime = \mathcal{H}_\mathrm{fut} \oplus \mathcal{H}_\mathrm{past},
\end{equation}
where $\mathcal{H}_\mathrm{fut}$ and $\mathcal{H}_\mathrm{past}$ are defined as isomorphic to $\mathcal{H}_\mathrm{local}$.  On this new Hilbert space, let the new past arrow $P^\prime$ and future arrow $F^\prime$ be defined by
\pagebreak[2]
\begin{eqnarray}
P^\prime \psi = P \psi:        & \psi \in \mathcal{H}_\mathrm{past}, \\
F^\prime \psi = 0 \phantom{P}: & \psi \in \mathcal{H}_\mathrm{past}, \\
P^\prime \psi = 0 \phantom{F}: & \psi \in \mathcal{H}_\mathrm{fut\phantom{t}}, \\
F^\prime \psi = F \psi:        & \psi \in \mathcal{H}_\mathrm{fut\phantom{t}},
\end{eqnarray}
so that $F^\prime$ and $P^\prime$ commute with one another.  This creates an additional single-vertex IER associated with the Hilbert space $\mathcal{H}_\mathrm{past}$, and an additional single-vertex FER associated with the Hilbert space 
$\mathcal{H}_\mathrm{fut}$.  

The evolution of these new single-vertex evolution regions is given by adding a new unitary transformation $Y(\mathcal{H}_\mathrm{past} \to \mathcal{H}_\mathrm{fut})$ to the dynamics, which encodes the fact that $\mathcal{H}_\mathrm{past}$ and $\mathcal{H}_\mathrm{fut}$ are really the same Hilbert space as each other.  Accordingly, $Y$ is simply the identity operator on $\mathcal{H}_\mathrm{local}$, interpreted as a unitary transformation taking vectors in $\mathcal{H}_\mathrm{past}$ to vectors in $\mathcal{H}_\mathrm{fut}$.  The other unitary operators associated with the other IER's and FER's (the ones that existed before the state doubling) remain unchanged.

Hence, by choosing a basis for the Hilbert space $\mathcal{H}^\prime$ compatible with the $F^\prime$ and $P^\prime$ operators, one recovers a DUCT which is in the same form as the DUCT's described in section \ref{kin}.  The laws of physics of this DUCT are determined by the full set of unitary transformations including the $Y$ matrix.

\subsection{A Bivalent DUCT}\label{biv}

A connected finite bivalent network is simply $N$ of vertices arranged into a circle.  If time continues forever in both directions, a bivalent DUCT will thus describe quantum spacetimes with spatial topology $S^1$.  (Presumably it is also possible to describe spatially infinite universes by some limiting process, but this will not be attempted here.)

The goal of this section is to make the simplest possible nontrivial DUCT for a bivalent network.  In order to count it as nontrivial, let us insist that the laws not conserve the total number of vertices $N$.  This requires processes in which the vertices can split or merge.
  
The simplest nontrivial model which permits this has a 3-dimensional state space at each point: Each vertex has a future arrow $f$ which points towards the vertex on the left (L), the vertex on the right (R), or else to the vertex itself (X).  Then an IER consists of either 1) a pair of adjacent vertices that point towards each other ($f_L \!\smile\! f_R$), 2) a vertex that points to itself ($f_X$), or 3) all of space, if the future arrows point in a circle.  Notice that it is impossible for two IER's to overlap, as required to avoid race conditions (cf. section \ref{motive}).

\begin{figure}[ht]
\centering
\includegraphics[width=.9\textwidth]{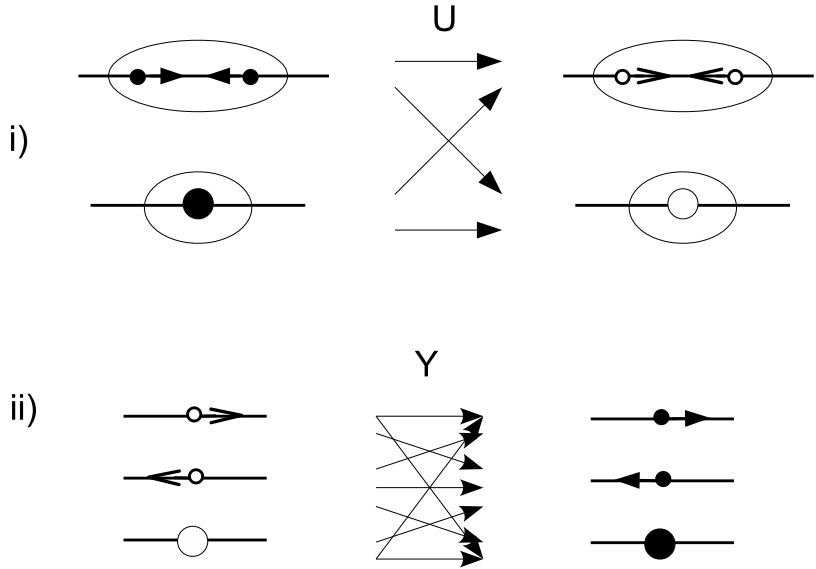}
\caption{ \footnotesize The two unitary transformations which describe the laws of physics of the bivalent DUCT: (i) the $U$ transformation converts future arrows to past arrows, in a way which can merge or split vertices.  Solid black represents future arrows, outlines represent past arrows.  Arrows that point to the vertex itself are shown as larger discs.  (ii) the $Y$ transformation converts past arrows back to future arrows.  It encodes the mixing between the past and future arrow bases.}\label{2dmoves}
\end{figure}

Similarly, in order to define past evolution, one needs to define a past arrow $p$ used to construct FER's.  This also requires a 3 dimensional space of states.  If the past and future arrows were independent degrees of freedom, this would require $3 \times 3 = 9$ different states per vertex.  This, however, is not necessary.  As discussed in section \ref{noncom}, it is possible for both arrows to be encoded in the same 3 dimensional Hilbert space, but using a different basis.  This is described by a $3 \times 3$ unitary matrix $Y$ which relates the $(f_L,\,f_R,\,f_X)$ basis to the $(p_L,\,p_R,\,p_X)$ basis (shown in Fig. \ref{2dmoves}).

In addition to the $Y$ matrix, one needs the unitary matrix $U$ which describes unitary evolution from FER's to IER's.  Since there are two kinds of IER's $(f_L \!\smile\! f_R,\,f_X)$ and likewise two kinds of FER's, this will be a 2x2 unitary matrix (also shown in Fig. \ref{2dmoves}).

The only situation in which $U$ and $Y$ are insufficient to evolve forwards, is when either all $N$ of the future arrows point left, or all of them point right.  In this case, the evolution is deadlocked unless such ``looped'' configurations are themselves regarded as IER's and evolved forwards in time to the corresponding FER's.  This adds one additional IER for each value of $N$ (assuming that the bivalent network is unoriented, so that there is no physical distinction between the state whose arrows point right and the state whose arrows point left).  A similar count applies to the FER's.  Consequently one needs to supply an additional infinite dimensional unitary matrix $U_\mathrm{loops}$ to specify the laws of physics.  However, additional principles might restrict the form of this matrix.  For example, one could require that $U_\mathrm{loops}$ must reduce to a rule which acts on each vertex independently.  This would reduce the information in $U_\mathrm{loops}$ to a single phase factor $e^{i \phi N}$.

Ignoring $U_\mathrm{loops}$, the choice of $Y$ and $U$ specifies the space of possible laws of physics for the DUCT.  A superficial count of degrees of freedom indicates $9 + 4 = 13$ different continuous parameters in the laws of physics, but this number must be adjusted to account for phase ambiguities and discrete symmetries.  There are 6 phase ambiguities coming from rotating the definitions of each of the states $f_L, f_R, f_X, p_L, p_R, p_X$ by a phase; consequently only $13 - 6 = 7$ degrees of freedom remain.

If the bivalent network is unoriented, the laws of physics must be invariant under parity ($P$) symmetry (which switches left and right).  Assuming that $p_X$ and $f_X$ are $P$-even, the matrix $Y$ acts on vectors with two $P$-even states and one $P$-odd state; consequently imposing parity reduces the $Y$ matrix to 5 degrees of freedom $2 \times 2 + 1 \times 1 = 5$ components.  Since all IER's and FER's are $P$-even, the $U$ matrix is unaffected.  However, since the ambiguous phases are also constrained by parity, their number is reduced from 6 to 4.  Consequently the number of constants becomes $9 - 4 = 5$.

One can also consider imposing time-reversal symmetry ($T$) (which swaps $p$ with $f$ while also taking the complex conjugate of the wavefunction).  Since $(U^*)^{-1} = U^T$, the effect of $T$ is to take the transposes of $U$ and $A$.  Hence $T$-invariance requires $U$ and $A$ to be symmetric.  However, it also relates 3 of the phase ambiguities to the other 3.  The effect of either $T$ or $PT$ invariance is to reduce the number of constants to $9 - 3 = 6$.

If one imposes both $P$ and $T$ symmetry, the number of constants is $7 - 2 = 5$.  This is the same as the number of constants when $P$ alone is imposed.  This is because if the DUCT satisfies $P$, the ambiguous phases can be chosen so that it also satisfies $T$ as well, as an accidental symmetry.

\subsection{A Trivalent DUCT}\label{triv}

\begin{figure}[ht]
\centering
\includegraphics[width=.7\textwidth]{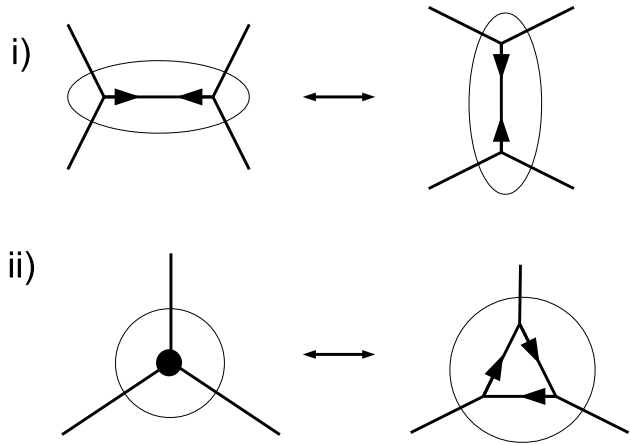}
\caption{ \footnotesize The two Pachner moves shown here can be used iteratively to turn any trivalent network into any other trivalent network.  Here the networks are shown labelled by the appropriate DUCT arrows to correspond to that move.  The arrows depicted are past arrows if occurring at the start of the move, future arrows if occurring at the end.  In move (i), a pair of vertices pointing at each other is replaced with a similar pair in a different orientation.  In move (ii), either a vertex pointing to itself is replaced with a triangle loop, or vice versa.}\label{pachner}
\end{figure}

Consider trivalent networks with unoriented edges and symmetric vertices.  Generalizing the previous section, a simple trivalent DUCT will be defined satisfying the following restrictions: a) there are only four states at each vertex, b) the network can be modified only by Pachner moves (see Fig. \ref{pachner}), and c) in the case of loops of four or more vertices, the evolution rule acts locally at each vertex.

To each vertex of the trivalent network, we assign a future arrow $f$ which either points either in one of the three directions, or to the vertex itself $(f_1, f_2, f_3, f_X)$.  Similarly, we define a past arrow, also with four states: $(p_1, p_2, p_3, p_X)$.  We need a unitary matrix $Y$ relating these four possibilities.  The laws of the DUCT must be symmetric under permutations of the 3 directions.  The four states subdivide into one 2-dimensional irrep of the permutation group $P_3$, and two copies of the 1-dimensional (trivial) irrep.  Consequently the matrix $Y$ has 5 parameters.

An IER is defined as a set of vertices whose future arrows point to one another in a loop.  There are an infinite set of these loops, so once again there is a danger of having infinitely many parameters.  In order to restrict the DUCT some limiting principle must be imposed; e.g. that any loop of $N \ne 3$ vertices evolves by means of a unitary operation which is independently applied to each vertex.  If this rule is selected, the evolution of such loops is given by a phase factor 
$U_N = e^{i \phi N}$, plus a single bit of information (whether the past arrow $p$ of the resulting FER is to go around the loop in the same direction as the future arrow $f$ did, or the opposite direction).\footnote{An interesting alternative would be to allow nearest neighbor interactions on the loop.  Unfortunately, there are many arbitrary choices which go into a DUCT.  Here I am trying to make the simplest possible model with nontrivial features.}

The case of a loop with $N = 3$ is special because it has the same number of outgoing edges as a single trivalent vertex.  Thus it belongs to the same boundary class as a IER or FER with just one vertex.  There are 3 kinds of IER's (FER's) in this boundary class: (a) the triangle loop with arrows going around clockwise, (b) the triangle loop with arrows going counterclockwise, and (c) a single vertex pointing to itself.\footnote{I have simplified the dynamics by ignoring the possibility of edges which connect two vertices in the same IER or FER.  When such internal edges appear, it is consistent to simply evolve as though the internal edge were two external edges, and just carry along the fact that they are connected to the final state.  It is also consistent, but more complicated, to expand the boundaries of the IER's and FER's to include edges which connect two of the vertices.  This would allow for examples of an IER $i$ with a nontrivial symmetry factor $|\mathrm{Aut}(i)| \ne 1$.} The transition amplitudes between these states are given by a unitary matrix $U_3$ (where the subscript counts the number of outgoing edges).  The 3 states split into two 1-dimensional (trivial) irrep of $P_3$, and one 1-dimensional (antisymmetric) irrep of $P_3$.  So $U_3$ has 5 parameters.

When two vertices point at each other along the same edge, there are four outgoing edges.  (This should be distinguished from an $N = 2$ loop, composed of two different internal edges.)  There are 3 different IER (FER) states corresponding to the 3 different ways in which the four edges can be associated.  These 3 states belong to one 1-dimensional (trivial) irrep of the permutation group $P_4$, and one 2-dimensional representation.  The unitary matrix $U_4$ relating the IER's to the FER's therefore only has 2 phase parameters.  (There would be more degrees of freedom if one were to allow the $N = 4$ loops to mix with these states, rather than giving them a separate evolution rule.)

Additionally, there are 4 phase ambiguities associated with rotating the phases of $(f_{1...3}, p_{1...3}, f_X, p_X)$.  The total number of parameters is therefore $13 - 4 = 9$.  If one imposes $T$-invariance there are still $11 - 2 = 9$ parameters, indicating the model has an accidental time-reversal symmetry.

\section{Prospects}\label{pros}

\paragraph{Summary of Results.} The DUCT framework is a promising way to explore discretized quantum gravity theories.  Like general relativity, it is defined in a way that makes no reference to any background space or time structures.  Causality is built into the theory at the fundamental level, so that each summed-over history has a well-defined causal structure.  Because different spatial regions are causally independent, the theory is also gauge-invariant under time lapses which can vary from place to place.

The absence of ``race conditions'' indicates that the algebra generating these gauge symmetries is closed, i.e. if there are two ways to evolve a slice forwards in time, it does not matter which occurs first.  The absence of ``deadlock'' ensures that time evolution is always possible from any slice.

The gauge-invariance of the theory was explicitly proven in section \ref{Degen} using the sum-over-histories formula of section \ref{rules}.  On the other hand, it has not yet been shown that the norm-squared of every state is positive, although this is plausible since one can prove that original states (those which have a beginning in time) have a positive norm (cf. section \ref{pos}).  Nor has it been shown that the inner product converges; for a realistic DUCT Abel summation is probably necessary and sufficient to obtain finite answers (cf. section \ref{resum} and Appendices).  

These conditions are sufficient for the DUCT to be consistent on the microscopic level, i.e. to have a well-defined Hilbert space, with the right number of states to describe causally propagating degrees of freedom without absolute space or time.  As argued in the Introduction, this is more than can be currently said for other approaches to quantum gravity, including causal sets, canonical Loop Quantum Gravity, spin-foams, and casual dynamical triangulations.  

\paragraph{A Continuum Limit?}  However, it is not enough for a discrete model of spacetime to be consistent on the microscopic quantum level.  In order to be a viable model of quantum gravity, it must also have a continuum limit which has 4 large spacetime dimensions, with dynamics approximately like that of classical general relativity (or some other gravitational theory similar enough to be experimentally viable).  It is therefore worthwhile to speculate about whether a DUCT-like model might, with the right values of the coupling constants, have such a continuum limit.

Like all other discrete quantum gravity models, it is unknown whether the DUCT spacetime behaves like a continuum theory at large distance scales.  The fact that the number of spacetime histories diverges at most exponentially is a promising feature; otherwise one would expect the large volume limit to be dominated by configurations far from any classical geometry, rather than by configurations close to some approximately classical solution.\footnote{This point is discussed in the literature on casual vs. noncausal dynamical triangulations \cite{AL98}.}

In the casual set approach to quantum gravity, the only geometrical structure is a partially ordered set of points.  When a causal set is sprinkled randomly on a background spacetime, it provides just enough information to approximately reconstruct the metric $g_{\mu\nu}$.  That is because the casual ordering of the points is a discrete analogue to the causal structure $\tilde{g}_{\mu\nu} = g_{\mu\nu} / (-g)^{D}$, while the number of points in a given spacetime volume is an analogue of the volume element $\sqrt{-g}$ \cite{csets}.  The DUCT histories include a causal set, plus additional structure (labelled edges and faces).  Therefore, if there are four large spacetime dimensions one may expect spacetime to be described by a metric plus additional fields.

However, there are arguments that any finite-valency graph structure necessarily breaks Lorentz invariance \cite{BHS09}.  One expects that the theory of quantum gravity should include an vacuum state, which should look approximately like Minkowski space in the continuum limit (for small values of the cosmological constant).  However, there is no way to embed such a structure in Minkowski space without picking out a frame, because the hyperboloid of unit timelike vectors has infinite volume \cite{BHS09}.  Admittedly, the relationship between the quantum histories and the classical histories may be more subtle than just an embedding, not only because of diffeomorphism invariance, but also because the classical history involves a sum over many different quantum histories.  However, I suspect that neither of these considerations is sufficient to evade the conclusion of Ref. \cite{BHS09}.

For this reason, I expect that the continuum limit---if it exists at all---will be an ``Einstein-Aether'' theory in which the metric $g_{ab}$ is dynamically coupled to a unit timelike vector \cite{JM01}.  Einstein-Aether theories break local Lorentz invariance, but preserve spacetime diffeomorphism-invariance.  As long as the matter fields we observe share a common lightcone, they are heavily constrained, but not currently eliminated, by observations; see Ref. \cite{jacobson08} for a recent review.  However, the renormalization group flow will typically lead to matter fields travelling at different maximum speeds, which is severely constrained by observation \cite{mattingly05}.

One way out is supersymmetry \cite{super}, but it is unclear how to supersymmetrize a DUCT.  Normally the anticommutator of the supersymmetry generators includes a \emph{continuous} time translation symmetry, but the DUCT has discrete time evolution.  It might be possible to find a limit of the coupling constants in which time evolution is effectively continuous, but this will not be attempted here.

Another way out would be if the DUCT approaches a Lorentz-invariant fixed-point, over many orders of magnitude in the renormalization group flow \cite{LVFP}.  In this case the discrete distance scale $L_D$ of the DUCT could be much shorter than the apparent Planck length $L_P$ as measured at large distances.  It might even be that one can take the $L_D \to 0$ limit: in this case the true theory of Nature would be an asymptotically safe Lorentz-invariant quantum gravity model \cite{NR06}, and the DUCT would merely be a UV regulator used as an intermediate step to define the theory.

Although this situation looks somewhat dire, it is worth pointing out that loop quantum gravity is in the same boat, since spin-foams also contain a finite-valence graph structure.  There are arguments that loop quantum gravity is manifestly Lorentz invariant \cite{RS11}, but I find them unpersuasive.  It is true that loop quantum gravity has an SL(2, C) gauge group, which is supposed to implement local Lorentz symmetry.\footnote{In the usual formulation of the canonical theory, this group is broken down to a subgroup SU(2), but there exists an equivalent formulation which is manifestly invariant under SL(2, C) \cite{RS11}.}  However, in order for this SL(2, C) gauge symmetry to count as Lorentz symmetry (rather than being an internal gauge group like that of QCD), there have to exist four large spatial dimensions whose tangent space transforms in the vector representation of the SL(2, C).  But then, if the orientation of the spin-foams picks out a special timelike direction in the tangent space, the local SL(2, C) symmetry will be spontaneously broken onto a smaller SU(2) subgroup.  Hence I expect loop quantum gravity must also violate Lorentz invariance.\footnote{However, if I am wrong and an SL(2, C) connection is sufficient to imply Lorentz invariance, then nothing prevents us from adding an SL(2, C) gauge connection to the DUCT graphs.}

\paragraph{Algebra of Diffeomorphisms}

Since the DUCT is defined in a background free way, i.e. without any reference to absolute space or time, one expects that a continuum limit (if it exists) will also have a background free formulation.  However, there are subtleties involving the constraint algebra and its associated conserved quantities which need to be analyzed carefully.

Microscopically, the DUCT is manifestly invariant under a discrete analogue of lapse and shift diffeomorphisms.  Invariance under shift diffeomorphisms is manifest by construction.  In section \ref{Degen} the inner product of the DUCT was explicitly shown to be degenerate under the discrete lapse gauge symmetry (something which has not yet been shown for spin foams \cite{LFprivate}).  

However, these discrete symmetries are probably not associated with any local conserved currents, since Noether's theorem only shows that \emph{continuous} symmetries give rise to conserved currents.  (In QFT, discrete symmetries typically give rise only to globally conserved charges.)  In the context of Regge calculus, the discretized Bianchi identity is not exact \cite{GKMM09}.  Presumably \emph{if} a discrete model of spacetime has a continuum limit, then the associated conserved currents exist only in that limit.

The absence of race conditions (i.e. no overlapping IER's or FER's) automatically implies that the algebra of discrete time evolution is manifestly closed, in the trivial sense that evolving one IER forwards commutes with evolving any other IER forwards.  However, one may wonder how this property would carry over to a continuum limit (if it exists).

In canonical general relativity, the algebra of diffeomorphisms is closed in a nontrivial way.  In particular, the Poisson bracket of the Hamiltonian constraint $H[N]$ with itself can be written in terms of the spatial diffeomorphism constraint $D[N^\mu]$ as follows \cite{dirac58}:
\begin{equation}\label{closed}
\{H(N),\,H(M)\} = \int d^{D-1}x D[(N \partial_\nu M - M \partial_\nu N) q^{\mu \nu}]
\end{equation}
where $N$ is the lapse, $N^{a}$ is the shift vector, and $q^{\mu \nu}$ is the spatial metric.  Intuitively, if the lapse function on an initial spatial slice is not constant, then the final slice is Lorentz-boosted relative to the initial slice, and therefore $H$ and $D_\mu$ mix.  This indicates that Eq. (\ref{closed}) encodes not only diffeomorphism invariance, but also local Lorentz invariance.

It is difficult to check the DUCT against Eq. (\ref{closed}), because in this article's formalism all states are invariant under spatial diffeomorphisms by construction, so any discrete analogue of $D_{\mu}$ will act trivially on the space of states.  However, the absence of any obvious discrete analogue of Eq. (\ref{closed}) raises legitimate worries about whether it can be recovered in the continuum limit.  This may be related to the Lorentz-violation issue described above.

\paragraph{Casual Horizons} Because DUCT's have a causal structure, it should also be possible to analyze the behavior of causal horizons in DUCT's.  It would be interesting to see whether these horizons obey the laws of thermodynamics.  In semiclassical gravity, the Generalized Second Law seems to depend on spacetime possessing local Lorentz invariance \cite{EFJW07}.  Since the discreteness of a DUCT is likely incompatible with local Lorentz invariance, it is not clear that these horizons should obey the laws of generalized thermodynamics.

\paragraph{Coupling to a Boundary.}  Another interesting consistency test involves coupling the DUCT to a spatial boundary, either at finite or infinite distance.  If one allows an absolute notion of `time' at the boundary (as usual in general relativity), then one could examine the spectrum of energies, and search for the lowest energy state.  This requires for time to be a continuous rather than a discrete parameter on the boundary; otherwise the energies would take values in U(1) rather than R.\footnote{If the boundary is at finite distance, this could be done by defining a unitary evolution operator $U$ at the boundary which is very close to the identity operator: $U = I + \mathcal{O}(\epsilon)$, and then taking the limit as $\epsilon \to 0$.  If the boundary is at infinite distance, it might also be possible to have an infinite redshift factor as in Anti-de Sitter space.}  If there is no lowest energy state, then the theory is presumably unstable.  If there is a lowest energy state, then it would be a good candidate for a vacuum-like state with a continuum limit.

There is a general argument by Marolf \cite{holo} that any generally covariant theory coupled to a boundary must have a holographic ``boundary unitarity'' property, in which the algebra of observables on the boundary is unchanged with time.  It would be interesting to check this argument in the context of a DUCT with boundary, especially since holographic principles 
were not used in any way to construct the theory.  Kinematically, the information in a DUCT is localized on the vertices of the network.  The dynamics are specified with quasi-local unitary transformations.  On the other hand, no diffeomorphism-invariant theory has truly local degrees of freedom, so the question is a nontrivial one.

\paragraph{Numerical Simulations.} To explore the above questions, it may be helpful to do computer simulations.  Unfortunately, since the number of histories with $t$ transitions can grow exponentially with $t$, a complete survey of histories is only possible for very small values of $t$, whereas a continuum limit would involve large values of $t$.  A DUCT is defined in Lorentzian signature (since $U$ can involve phases), so Monte Carlo simulations are impossible.\footnote{Unless one is only interested in whether the sum over histories is absolutely convergent, in which case only the absolute values of the transition amplitudes matter.}

Nevertheless, it is possible to analytically continue to a theory which---like a Wick rotated QFT---involves only positive real amplitudes.   (I will call this theory ``Euclidean'' without regard to whether it corresponds in any way to Riemannian geometries.)  One loses some key physical properties of the inner product, but this does not matter so long as one is only interested in the Euclidean model as a diagnostic for some property of the Lorentzian theory.

Here is one way to do this:  Each unitary matrix $U$ can always be written in the form $E = Je^{i \lambda S} = U$, where $J$ replaces each IER with the time-reversed FER (without complex conjugating), $S$ is a hermitian operator defined on the vector space of IER's, and $\lambda = 1$ initially.\footnote{This construction is not unique since one can add any multiple of $2\pi$ to the eigenvalues of $S$ without changing the dynamics.}   Changing $\lambda$ to another real value deforms the coupling constants of the DUCT to another unitary matrix $U^\prime$.  However, if one analytically continues to imaginary $\lambda$, then every forward-evolving history has a positive real transition amplitude.  If $S > 0$, and $\lambda$ is positive imaginary, this transition amplitude is always less than 1.

Backwards-evolving histories pose a bit of a challenge.  If one evolves backwards in time using $E^{-1} = e^{-i \lambda S}J^{-1}$, then the inner product is no longer hermitian, and also there may be problems with convergence for positive imaginary values of lambda.  Alternatively, one can evolve backwards using $E^\dagger = e^{-i \lambda^{*} S}J^{-1}$, so that the inner product is hermitian, but no longer analytic in $\lambda$.  (One would then have to analytically continue the forwards and backwards parts of the inner product separately.)  In neither case is the inner product gauge-degenerate, since this requires $E$ to be unitary (cf. section \ref{Degen}).  However, this Euclidean inner product should be numerically tractable, and it may reveal physically interesting properties of the Lorentzian signature theory.\footnote{It is worth pointing out that there is no analogue of the Hartle-Hawking state in this Euclidean sum over histories, since it does not include any processes which change the number of connected components of space.}

\subsubsection*{Acknowledgements}
\small
I am grateful for comments from William Donnelly, Laurent Freidel, Don Marolf, Lee Smolin, Ted Jacobson, and anonymous referees.  Supported by NSF grants PHY-0601800, PHY-0903572, the Maryland Center for Fundamental Physics, the Simons Foundation, and the Perimeter Institute.
\normalsize
\nopagebreak
\appendix
\section{Divergences are at most Exponential}\label{bound}
In this appendix it will be shown that divergences in Eq. (\ref{DIP2}), the sum over histories, cannot be more severe than exponential in $t$, the number of transitions.

\subsection{Combinatoric Bound on the Sum over Histories}

In order to analyze the convergence properties of the DUCT, let us make the simplifying assumption that each unitary transition matrix in the theory has at most $M$ possible initial or final states, where $M$ is a single finite constant.\footnote{The particular DUCT's defined in section \ref{exam} can be considered to satisfy this property.  The maximum dimension of any nontrivial unitary matrix is given by $Y$, which is $3 \times 3$ for the bivalent DUCT, and $4 \times 4$ for the trivalent DUCT.  Although $U_\mathrm{loops}$ (in the bivalent case) and $U_N$ for $N > 5$ (in the trivalent case) have an unbounded number of possible states, because these unitary matrices were required to be a phase factor $e^{i \phi N}$, one can break up these matrices into an infinite number of $1 \times 1$ dimensional blocks.  Thus, for purposes of the bound below, the bivalent DUCT has $M = 3$ while the trivalent DUCT has $M = 4$.}  It can then be shown that for a fixed initial slice $A$ and final slice $B$, the sum over histories cannot diverge any worse than exponentially in $t$, the total number of transitions in the history.

In order to specify a history beginning at $h_0$ it is sufficient to determine the following data: a) whether each of the initial vertices in $A$ is evolved forwards or backwards in time (or remains unevolved), b) for each IER which appears in the course of forwards evolution, whether that IER is evolved forwards in time or remains in the final slice, c) similarly, for each FER which appears in the course of backwards evolution, whether it is evolved backwards or not, and d) for each evolved IER or FER, which of at most $M$ possible transitions occurs at that site.  (This is an overestimate since there are certain constraints on this data, such as the requirement that the final slice be $B$.)

The possibilities for (a) is bounded by $3^{V(A)}$, where $V(A)$ is the total number of vertices of the initial slice.  Since $A$ is fixed, this is a constant with respect to $t$.  The possibilities for (b) and (c) are bounded by $2^{t + V(B)}$, where $V(B)$ is the total number of vertices in the final slice.  This is because each evolution region in the history is either evolved (which uses up one of the transitions $t$, or else remains on the final slice (and thus occupies at least one vertex there).  Finally, (d) is bounded by $M^t$.  Consequently, the number of possible histories used to calculate $\langle B | A \rangle$ grows at most exponentially with $t$.

Ignoring symmetry factors, each of these histories has a sub-unit absolute amplitude $|\mathcal{A}(h)| \le 1$, since the amplitude is the product of elements of unitary matrices.  This means that the total amplitude from all histories is at worst exponentially divergent in $t$.

\subsection{Symmetry Factors}

This conclusion does not change when symmetry factors are taken into account.  Here I will indicate the essential reason for this without providing a full rigorous proof.  

The potential problem is that the inclusion of the Eq. (\ref{T}) shift-automorphism symmetry factor $\sqrt{ |\mathrm{Aut}(a)| \cdot |\mathrm{Aut}(i)| }$ can cause the absolute amplitude of a single history to exceed one.  However, the additional symmetry of the FER or IER always leads to additional consequences that correspondingly lower the sum over histories.  For any history $h$ containing a FER $a$ which is symmetric under the shift automorphism $q$, one can define a transformed history $h^\prime$ which is just the same as $h$ except for the insertion of the shift automorphism $q$ right after $a$ is produced.  Then either (A) $q$ extends to an automorphism of the whole history $h$ (not counting symmetries of the initial and final states), or (B) $q$ identifies two histories which would otherwise be distinct.

In case (A), $q$ acts on the history $h$ in a compact region, whose last transition must be an IER symmetric under $q$, just as its first transition was a FER symmetric under $q$.  However, there is a compensating factor of $1/|\mathrm{Aut}(h)|$ coming from Eq. (\ref{amp}) which cancels out the IER and FER symmetry factors, so that $|\mathcal{A}(h)| \le 1$.

In case (B), the symmetry leads to the identification of two histories which would otherwise be counted as distinct for purposes of the bound on the number of histories.  Thus the ability to have histories with $|\mathcal{A}(h)| > 1$ is compensated for by the fact that there are correspondingly fewer histories included in the count.  

Thus in no case can the symmetry factors increase the maximum sum of the absolute amplitudes of all histories.

\section{Cyclic Cosmologies}\label{cycle}

It is possible in principle for a spacetime to have a periodic history, in the sense that after a certain amount of time, the configuration of space repeats its initial starting point.  In these situations the inner product displays interesting quantum interference effects, even when the dynamics of the DUCT are otherwise fully classical (i.e. deterministic).

\subsection{Classical Limit}

There is an interesting limit of DUCT's in which they show nearly classical behavior.  If one chooses each unitary transformation so that, in the arrow basis, the absolute values of all of the matrix entries are either 0 or 1, then the evolution of spacetime proceeds by deterministic rules.  This provides a limit in which the theory essentially reduces to a \emph{classical} discrete causal background-free geometry.  Obviously this is very different from the classical limit---which may or may not exist---in which the theory reduces to a spacetime continuum theory such as semiclassical general relativity.  (The possible classical limits of the DUCT's described in sections \ref{biv} and \ref{triv} are too simple to have interesting classical limits, but the addition of extra information can lead to nontrivial theories.)

However, even in this ``classical'' limit, the inner product (\ref{DIP2}) still provides for quantum interference effects for classical spacetimes whose evolution is periodic in time.  In this case, the sum over histories is not absolutely convergent, but is instead given by the sum
\begin{equation}\label{phases}
\langle \Psi | \Psi \rangle = \sum_{n = -\infty}^{+\infty} \chi^n,
\end{equation}
where $\chi$ is the phase contribution from a single period of the history.  If $\chi = 1$, the inner product becomes $+\infty$, while if $\chi$ is any other unit phase, the infinite series can be Abel resummed to $0$ (since it corresponds to a state vector which is equal to itself times a phase).

However, this extreme behavior is only seen if the probability of eventually returning to the same initial configuration is 1.  The sum becomes finite and nonzero if one multiplies by a real damping factor $c < 1$:
\begin{equation}\label{damp}
\langle \Psi | \Psi \rangle = \sum_{n = -\infty}^{+\infty} \chi^n c^{|n|} =
\frac{1 - c^2}{|1 - \chi c|^2}.
\end{equation}
In section \ref{resum}, the coupling constant $c$ was introduced as a technique for resumming nonconvergent path integrals.  In that section, $c$ was interpreted as an unphysical parameter which must be analytically continued to $c = 1$ in order to obtain a physical result.  However, it is also possible to reconcile $c < 1$ with unitarity, by postulating the existence of a process with amplitude $\sqrt{1 - c^2}$ for the system to tunnel out of the periodic cycle into another state with different behavior (such as eternal expansion or a final singularity).  This indicates that even a very small quantum probability to tunnel out of the periodic cycle into a different noncyclic asymptotic behavior, can be sufficient to make the inner product converge (so long as no new divergences are introduced by this new process).

Note that the inner product Eq. (\ref{damp}) is always positive for $c < 1$.  This is a special case of the positivity result of section \ref{pos}.  Because there is a nonzero amplitude at each time-step for each state to leave the cyclic history when evolved backwards (or forwards) in time, and because these additional states can be taken to be initial (or terminal) respectively, the projection operator $P$ onto initial (or terminal) states equals the identity.

\subsection{Quantum Periodicity}\label{quant}

Even away from the classical limit, similar behavior should appear whenever periodicity is inevitable, due to the system having only a finite number of states to explore.  For example, if one were to chose the unitary evolution matrices $U$ such that the total number of vertices were conserved (while preventing deadlock).  In such cases, for generic choices of the couplings there will be \emph{no} physical states, since all states will eventually evolve to themselves times a nontrivial phase.

As a particularly simple example, consider a DUCT with just one vertex, equipped with a finite-dimensional Hilbert space $\mathcal{H}$, and a unitary evolution operator $U(\mathcal{H} \to \mathcal{H})$.  One can choose a basis in which $U$ is diagonal.  By performing the sum over histories in this basis, one finds that the inner product for each basis state is given by Eq. (\ref {phases}).  This indicates that for generic choices of $U$, there are no states in the physical Hilbert space!\footnote{Recall that in the picture used here, the physical state space is obtained by modding out all vectors with zero inner product.  These states are dual to the states which satisfy the (discrete analogue of the) Hamiltonian constraint.  In this case, generically, there are no solutions to the (discrete analogue of the) Hamiltonian constraint.  This should not be a surprise.  Normally a nontrivial theory with a Hamiltonian constraint needs to have at least one degree of freedom whose dynamical range is unbounded.  Otherwise the Hamiltonian operator will have a discrete spectrum, and generically there will be no solutions to the constraint.}

Let us regulate this periodic theory using the damping factor $c < 1$.\footnote{One can simulate this damping within the rules of a DUCT by tripling the size of the Hilbert space $\mathcal{H}$, adding one new original IER state and one new terminal FER state for each old state.  One then chooses a unitary operator $U(\mathcal{H} \oplus \mathcal{H}_\mathrm{orig} \to \mathcal{H} \oplus \mathcal{H}_\mathrm{term})$ with the matrix elements $[cU,\,\sqrt{1 - c^2}U;\,\sqrt{1 - c^2}U,\,-U]$.} If one evaluates this theory in the basis in which $U$ is diagonal, one can find the norm of each basis state using Eq. (\ref{damp}).  Since there are no histories which mix the basis states, different basis states are orthogonal, and the inner product is
\begin{equation}\label{delt}
\langle B | A \rangle = \delta_{AB} \frac{1 - c^2}{|1 - \chi_A c|^2},
\end{equation}
where $\chi_A$ is the phase of $U$ evaluated in the basis state $A$.

What if the inner product is evaluated in a different basis, such that $U$ is not diagonal?  One would like the answer to be just Eq. (\ref{delt}) rotated into the new basis.  In fact, the sum over histories is no longer absolutely convergent in the new basis.  Recall that in order for a series to be absolutely convergent, the sum of the absolute values of the squares must converge.  But in the case of a nondiagonal unitary operator,
\begin{equation}
\sum_i | U_{ij} | > \sum_i | U_{ij} |^2 = 1,
\end{equation}
which implies that the absolute sum over all amplitudes grows exponentially with time.  This means that the sum over histories is not absolutely convergent even when $c$ is slightly less than 1.  Nevertheless, by adding up the histories in the right order (performing the sum over all histories with a given $t$ before summing over $t$), one gets the right answer.  (Hence one also obtains Eq. (\ref{delt}) using the stronger method of Abel summation.)

\end{document}